\documentclass[11pt,a4paper]{article}

\usepackage[T1]{fontenc}
\usepackage[utf8]{inputenc}
\usepackage{lmodern}
\usepackage[margin=2.6cm,marginparwidth=2.2cm]{geometry}
\usepackage{amsmath,amssymb,amsthm}
\usepackage{mathtools}
\usepackage{enumitem}
\usepackage{booktabs}
\usepackage{tikz}
\usetikzlibrary{arrows.meta,calc}
\usepackage[colorinlistoftodos,textsize=small]{todonotes}
\usepackage{aliascnt}
\usepackage[hidelinks]{hyperref}
\usepackage[capitalise,nameinlink]{cleveref}

\hypersetup{
  pdftitle={Directed k Disjoint Shortest Paths Is Polynomial for Every Fixed k},
  pdfauthor={Saeed Amiri},
  pdfsubject={},
  pdfkeywords={disjoint shortest paths, directed graphs, polynomial-time algorithm, dynamic programming},
  pdfcreator={},
}

\theoremstyle{plain}
\newtheorem{theorem}{Theorem}[section]
\newaliascnt{lemma}{theorem}
\newtheorem{lemma}[lemma]{Lemma}
\aliascntresetthe{lemma}
\newaliascnt{proposition}{theorem}
\newtheorem{proposition}[proposition]{Proposition}
\aliascntresetthe{proposition}
\newaliascnt{corollary}{theorem}
\newtheorem{corollary}[corollary]{Corollary}
\aliascntresetthe{corollary}
\newcommand{\restatedname}{}
\newtheorem*{restatedthm}{\restatedname}
\newenvironment{restated}[1]{\renewcommand{\restatedname}{\Cref*{#1}}\begin{restatedthm}}{\end{restatedthm}}
\theoremstyle{definition}
\newaliascnt{definition}{theorem}
\newtheorem{definition}[definition]{Definition}
\aliascntresetthe{definition}
\theoremstyle{remark}
\newaliascnt{remark}{theorem}
\newtheorem{remark}[remark]{Remark}
\aliascntresetthe{remark}

\newcommand{\R}{\mathbb{R}}
\newcommand{\dist}{d}
\newcommand{\pt}{\pi}
\newcommand{\one}{\mathbf{1}}
\newcommand{\X}[1]{X_{#1}}
\newcommand{\ext}{\operatorname{ext}}
\newcommand{\Pk}{P(k)}
\newcommand{\Fam}{\mathcal{U}}
\newcommand{\Wzero}{\textup{(W0)}}
\newcommand{\Wone}{\textup{(W1)}}
\newcommand{\Wtwo}{\textup{(W2)}}
\newcommand{\Mone}{\textup{(M1)}}
\newcommand{\Mtwo}{\textup{(M2)}}
\newcommand{\Tab}{\mathsf{T}}
\newcommand{\Sem}{\mathsf{S}}

\title{Directed $k$ Disjoint Shortest Paths Is Polynomial\\ for Every Fixed $k$}
\author{Saeed Amiri\\ \href{mailto:saeed.amiri.akh@gmail.com}{\texttt{saeed.amiri.akh@gmail.com}}}
\date{}

\begin{document}
\maketitle

\begin{abstract}
The \emph{$k$ disjoint shortest paths problem} ($k$-DSPP) asks for pairwise disjoint paths
between $k$ terminal pairs, each a shortest path between its own terminals; the paths are
required to be vertex-disjoint or, in the edge-disjoint version, edge-disjoint.  Eilam-Tzoreff
introduced the problem in 1998 and left its complexity for a fixed number of pairs open.  For
general graphs with arbitrary positive lengths, the case of three or more pairs remained open
for many years.  Earlier algorithms for general undirected graphs with unit lengths had
running-time exponents doubly exponential or factorial in~$k$.  We show that for every
fixed~$k$ both versions of $k$-DSPP are solvable in time $\smash[t]{n^{O(k^2)}}$ on directed graphs
with arbitrary positive lengths, and hence on undirected graphs.  The dependence of the exponent
on~$k$ cannot be removed unless FPT${}={}$W[1].
\end{abstract}


\section{Introduction}\label{sec:intro}

Let $G=(V,A)$ be a directed graph with arc lengths $\ell\colon A\to\R_{>0}$, and let
$(s_1,t_1),\dots,(s_k,t_k)$ be terminal pairs.  The \emph{$k$ disjoint shortest paths problem}
($k$-DSPP) asks whether there are pairwise vertex-disjoint paths $P_1,\dots,P_k$ such that each
$P_i$ is a shortest $s_i$--$t_i$ path \emph{of $G$}, that is, $\ell(P_i)=\dist(s_i,t_i)$, where
$\dist$ is the distance in $G$.  This is the vertex-disjoint version; in the \emph{arc-disjoint}
version, and in the \emph{edge-disjoint} version for undirected graphs, the paths are required to
be pairwise arc-disjoint or edge-disjoint instead.

Shortest is meant with respect to the whole graph, not among
disjoint solutions, and every path must be individually shortest; the problem is therefore
different from the min-sum disjoint paths problem, which minimizes the total length of $k$
disjoint paths.  That problem goes back to Suurballe~\cite{Suurballe74}, who solved it for paths
between a single pair of terminals; for later work, including several terminal pairs,
see~\cite{SuurballeTarjan84,KobayashiSommer10,CdVSchrijver11,BjorklundHusfeldt19,MMPS24}.  Ties are allowed throughout: a pair may have many shortest paths, and an arc may
lie on shortest paths of several pairs.

\begin{theorem}[main theorem]\label{thm:main}
  For every fixed $k\ge 2$ there is an algorithm with the following input and output.  The
  input is a directed graph $G$ with $n$ vertices, $m$ arcs and strictly positive arc lengths, and
  $k$ terminal pairs with $2k$ distinct terminals.  The algorithm decides whether $G$ has pairwise
  vertex-disjoint paths $P_1,\dots,P_k$ such that each $P_i$ is a shortest $s_i$--$t_i$ path in
  $G$, and finds such paths if they exist.  Its running time is
  \[
     O\bigl(m+ n\log n\bigr)+ f(k)\, n^{6k^2-4k-1}
  \]
  for a function $f$ of $k$ alone; in particular it is $\smash[t]{n^{O(k^2)}}$.  The algorithm only adds
  and compares lengths, so for rational lengths it runs in time polynomial in the input size.
\end{theorem}

An undirected graph with
positive edge lengths becomes a directed one by replacing every edge by two opposite arcs of the
same length; distances, shortest paths and vertex-disjointness are unchanged.  Hence:

\begin{corollary}\label{cor:undirected}
  For every fixed $k$, $k$-DSPP in undirected graphs with strictly positive edge lengths is
  solvable in time $\smash[t]{n^{O(k^2)}}$.
\end{corollary}

The edge-disjoint versions reduce to the vertex-disjoint one on directed graphs through line
graphs, with the shortest-path condition preserved;~\cref{sec:remarks} gives the reductions and
the proof of the following corollary.

\newcommand{\coredgedisjoint}{%
  For every fixed $k$ and strictly positive lengths, the following problems are solvable in time
  $O(m)+\smash[t]{n^{O(k^2)}}$, where $m$ is the number of arcs or edges:
  \begin{enumerate}[label=(\roman*)]
    \item the arc-disjoint version of $k$-DSPP in directed graphs;
    \item the edge-disjoint version of $k$-DSPP in undirected graphs.
  \end{enumerate}
  This holds also with parallel arcs or edges and when terminals of different pairs coincide
  (with $s_i \ne t_i$ for every~$i$).}

\begin{corollary}\label{cor:edgedisjoint}
  \coredgedisjoint
\end{corollary}

The same algorithm decides whether a path, a cycle or a waypoint route through a fixed number
of prescribed vertices can be as short as the distances between these vertices allow
(\cref{cor:waypoints} in~\cref{sec:remarks}).  In digraphs, finding a path through one prescribed
vertex or a cycle through two is NP-hard~\cite{FHW80}.

\Cref{thm:main} gives a polynomial-time algorithm for directed $k$-DSPP with arbitrary positive
lengths and a running-time exponent quadratic in~$k$.  For undirected graphs, the same bound
holds by \cref{cor:undirected}.  For general undirected graphs with unit lengths, the earlier
bounds were $\smash[t]{\cramped{n^{O(k^{5^k})}}}$~\cite{Lochet21} and
$n^{O(k\cdot k!)}$~\cite{BNRZ21}.

This is not a claim of speed for small~$k$.  The theorem of~\cite{BNRZ21} states the exponent
$16k\cdot k!+k+1$, which is $292$ for $k=3$ and $1541$ for $k=4$, against our $41$ and $79$; but
neither analysis is optimized, and for $k=2$ far faster algorithms are known~\cite{AVW24,BNRZ21,CKS26}.

\paragraph{Independent work.}  Cao, Qi and Wang~\cite{CQW26FixedK} posted a preprint on
25 November 2026 giving a deterministic algorithm for directed vertex-disjoint $k$-DSPP with
positive lengths.  Their Theorem~5.10 bounds the running time of a sweep with bounded memory
by $O_k(n^{g(k)+2})$, where $g(k)=2^{O(2^k k\log k)}$.  The quadratic exponent of our
guillotine-cut dynamic program answers the first part of their Question~11.1.

The work reported here was developed independently.  A complete draft with the quadratic bound
was already in place before their preprint appeared.  The author deferred posting while revising
and simplifying the manuscript.

\paragraph{From cycles to acyclic pieces.}  The shortest $s_i$--$t_i$ paths of one pair form an
acyclic subgraph $D_i$ of $G$.  If the union of $D_1,\dots,D_k$ were acyclic as well, the pebbling
algorithm of Fortune, Hopcroft and Wyllie~\cite{FHW80} would solve the problem: it moves all
paths forward along one common topological order.  In general the union has directed cycles; the
pairs disagree about which direction is forward, and this is the whole difficulty.

We make the
disagreement measurable.  For every pair $i$ we use a potential that measures how far a vertex is
on the way from $s_i$ to $t_i$, and we place every vertex at the point of $\R^k$ given by its $k$
potentials.  In this picture a shortest $s_i$--$t_i$ path never turns back: along it the
potential of pair $i$ grows at least as fast as every other potential, so the path moves steadily
within a fixed range of directions.  On vertices that lie at one point all pairs agree: an arc
of any $D_i$ between two such vertices raises all $k$ potentials by the same amount, so the union
of the $D_i$ restricted to one point is acyclic and the pebbling game applies there.  The cycles
run between different points, and we deal with them by cutting the space apart.

The core of the proof is a structural statement about any $k$ shortest paths, disjoint or not
(\cref{thm:hierarchy}).  Since none of them turns back, the space can be cut recursively into
boxes by \emph{guillotine cuts},\footnote{A \emph{guillotine cut} splits a region into two parts by a single cut that
runs all the way across it.  The term comes from the two-dimensional cutting stock problem,
where every cut runs from one edge of the sheet to the opposite edge~\cite{GilmoreGomory65}.  In algorithm design,
recursive guillotine subdivisions solved by dynamic programming give approximation schemes
for geometric problems such as the Euclidean TSP~\cite{Mitchell99}.} in the manner of a kd-tree~\cite{Bentley75}.
Only $O(k)$ arcs of the paths cross each cut, and every box contains only $O(k^2)$ pieces
of them, however large the graph is.

A dynamic program over the boxes then decides the problem
from the bottom up.  At a single point it runs the pebbling game; a larger box it splits into two
smaller ones by a cut, guessing the few arcs that cross the cut and the pieces they connect.  Two
pieces in different boxes never share a vertex, so disjointness only has to be checked inside the
boxes, among a few pieces, and the few arcs across each cut keep the number of table entries
polynomial for fixed~$k$.

\subsection{Previous work}\label{sec:known}

The \emph{$k$ disjoint paths problem} asks for pairwise
vertex-disjoint paths between $k$ terminal pairs, with no condition on their lengths.  It is
NP-complete when $k$ is part of the input~\cite{Karp75}.  For fixed $k$ the undirected and the
directed problem behave very differently.

In undirected graphs the case $k=2$ was solved by
Seymour~\cite{Seymour80}, Thomassen~\cite{Thomassen80} and Shiloach~\cite{Shiloach80}, and
Robertson and Seymour~\cite{RS95} gave an algorithm with running time $f(k)\,n^3$ for every
fixed $k$, a cornerstone of their Graph Minors project.  It is based on the \emph{irrelevant
vertex technique}.  As long as the treewidth of the graph is large, one finds a vertex whose
deletion does not change the answer, typically the center of a large flat wall, and deletes
it~\cite{RS95,RS12}.  Once the treewidth is bounded by a function of $k$, one solves the problem
by dynamic programming over a tree decomposition.  The running time was improved to $f(k)\,n^2$ by
Kawarabayashi, Kobayashi and Reed~\cite{KKR12} and to almost linear by Korhonen, Pilipczuk and
Stamoulis~\cite{KPS24}.

In directed graphs the problem is NP-complete already for
$k=2$~\cite{FHW80}.  On directed acyclic graphs it is solvable in time $n^{O(k)}$ by the pebbling
algorithm of Fortune, Hopcroft and Wyllie~\cite{FHW80}, and the edge-disjoint version is W[1]-hard
on them parameterized by $k$~\cite{Slivkins10}.  On planar digraphs it is solvable in time
$n^{O(k)}$ by Schrijver's algorithm~\cite{Schrijver94} and is fixed-parameter tractable in
$k$~\cite{CMPP13}.

\paragraph{Disjoint shortest paths.}  Eilam-Tzoreff~\cite{EilamTzoreff98} introduced the
disjoint shortest paths problem and showed that it is NP-hard when $k$ is part of the input.
She gave a polynomial-time algorithm for two pairs in undirected graphs with positive lengths.
She also showed that the condition must hold for every pair: finding two disjoint paths of which only the
first is required to be shortest is NP-complete, directed or undirected, even with unit
lengths.  She closed with the question that this paper answers: ``the complexity of the
undirected kDSP problem for fixed $k\ge 3$ and the directed kDSP problem for fixed $k\ge 2$ is
left open''~\cite{EilamTzoreff98}.

The irrelevant vertex technique does not help here.  Its rerouting arguments replace parts of
the solution paths by other paths, which need not be shortest, and deleting a vertex can change
the distances between the terminals; in this problem essentially every vertex on a shortest
path may matter.  The shortness condition gives something in return: every solution path moves
steadily away from its source, and the algorithms for disjoint shortest paths turn this
monotone structure into a dynamic program.

When the shortest-path subgraphs of all pairs fit
into one acyclic graph, the pebbling algorithm of~\cite{FHW80} applies to them
directly~\cite{BercziKobayashi17}.  Lochet~\cite{Lochet21} gave the first technique for general
undirected graphs.  Guessing where the paths of different pairs meet, he cuts them into a
bounded number of subpaths, such that subpaths of different pairs can no longer intersect.  He
then runs an adaptation of the pebbling algorithm on them.  Bentert et~al.~\cite{BNRZ21} reached a much
smaller exponent by a geometric analysis of how shortest paths of different pairs meet.  Our
potentials are a third way to tame the cycles; they work for directed graphs and arbitrary
positive lengths.

The shortness condition also changes the complexity: directed
$2$-DSPP with positive lengths is polynomial~\cite{BercziKobayashi17}, while the directed two
disjoint paths problem is NP-complete~\cite{FHW80}.

\paragraph{Undirected graphs.}  For two pairs
in undirected graphs polynomial-time algorithms are known also for non-negative lengths~\cite{GKW19,KobayashiSako19}.  For every fixed $k$, Lochet~\cite{Lochet21} gave a polynomial-time
algorithm for undirected graphs with unit lengths, with running time $\smash[t]{\cramped{n^{O(k^{5^k})}}}$;
Bentert, Nichterlein, Renken and Zschoche~\cite{BNRZ21} improved this to
$O(k\,n^{16k\cdot k!+k+1})$ by a geometric argument, and showed that the problem is W[1]-hard
parameterized by $k$.  They name the reduction of the
factor $k!$ in the exponent to a polynomial in $k$ as a challenge.  Akmal, Vassilevska Williams
and Wein~\cite{AVW24} detect two disjoint shortest paths in linear time in weighted undirected
graphs and in DAGs, and remark that an extension of the fixed-$k$ algorithms to weighted graphs
does not appear to be known.  For undirected planar graphs the problem is polynomial for fixed
$k$~\cite{BercziKobayashi17}, and with positive lengths it is fixed-parameter tractable in $k$~\cite{PSW26}.

\paragraph{Directed graphs.}  B\'erczi and Kobayashi~\cite{BercziKobayashi17} gave a
polynomial-time algorithm for directed $2$-DSPP (vertex- and edge-disjoint) with positive lengths;
it reverses the shortest-path DAG of one pair and applies a pebbling argument to the union.
They extend it to a fixed number of internally vertex-disjoint (or edge-disjoint) shortest paths
between only two terminal pairs.  Faster algorithms for directed $2$-DSPP were given by Choudhary, Kumar and Saggi~\cite{CKS26}
and, in a recent preprint, by Cao, Qi and Wang~\cite{CQW26}.

If zero-length arcs are allowed,
directed $2$-DSPP is NP-hard: with all lengths $0$ every path is shortest, and the problem
becomes the directed two disjoint paths problem, which is NP-hard by Fortune, Hopcroft and
Wyllie~\cite{FHW80} (see~\cite{BercziKobayashi17}).

On directed acyclic graphs $k$-DSPP is
polynomial for every fixed $k$ by the pebbling algorithm of~\cite{FHW80} applied to the
shortest-path subgraphs~\cite{BercziKobayashi17}.  On planar digraphs the vertex-disjoint
problem is polynomial for every fixed $k$ by Schrijver's algorithm~\cite{Schrijver94} applied to
the same subgraphs~\cite{BercziKobayashi17}.

Akmal, Vassilevska Williams and Wein~\cite{AVW24} reduce $k$-Clique to vertex-disjoint $k$-DSPP
on unweighted DAGs.  Bentert, Fomin and Golovach~\cite{BFG25} independently give essentially the
same reduction.  Amiri and Wargalla~\cite{AW21}
consider a version with congestion~$c$, in which every vertex may lie on up to $c$ of the paths,
and show W[1]-hardness on DAGs with respect to $k-c$ for every constant~$c$.  For $c=1$ this is
the vertex-disjoint problem on DAGs, parameterized by~$k$, and the hardness holds already with
unit lengths.  They also show W[1]-hardness of the edge-disjoint version on planar
DAGs.

\paragraph{Eilam-Tzoreff's question.}  For general graphs with arbitrary positive lengths, the
case $k=2$ was settled in the directed~\cite{BercziKobayashi17} and
undirected~\cite{EilamTzoreff98} settings.  The directed case for fixed $k\ge 3$ was raised
repeatedly~\cite{AW21,AkmalWein23,AVW24,BFG25,CKS26} and is now settled by the two independent
algorithms.

\subsection{Organization}

\Cref{sec:sketch} explains the whole argument informally.  \Cref{sec:prelim} sets up the
shortest-path DAGs, the potentials and the cells.  \Cref{sec:hierarchy} proves the structural
theorem (\cref{thm:hierarchy}).  \Cref{sec:algorithm} gives the dynamic program, proves that it
is correct, counts its running time and proves~\cref{thm:main}.  \Cref{sec:remarks} collects remarks
(zero-length arcs, what the proof uses, arc- and edge-disjoint paths, lower bounds) and open problems.

\section{The argument in brief}\label{sec:sketch}

This section explains the proof informally;~\cref{sec:prelim,sec:hierarchy,sec:algorithm} give
the details.

On acyclic graphs Fortune, Hopcroft and Wyllie~\cite{FHW80} solved the
disjoint paths problem by a dynamic program: their pebbling algorithm sweeps along a topological
order, and the part already swept meets the rest only in the current positions of the $k$
pebbles.  The acyclic graph supplies this small interface for free.

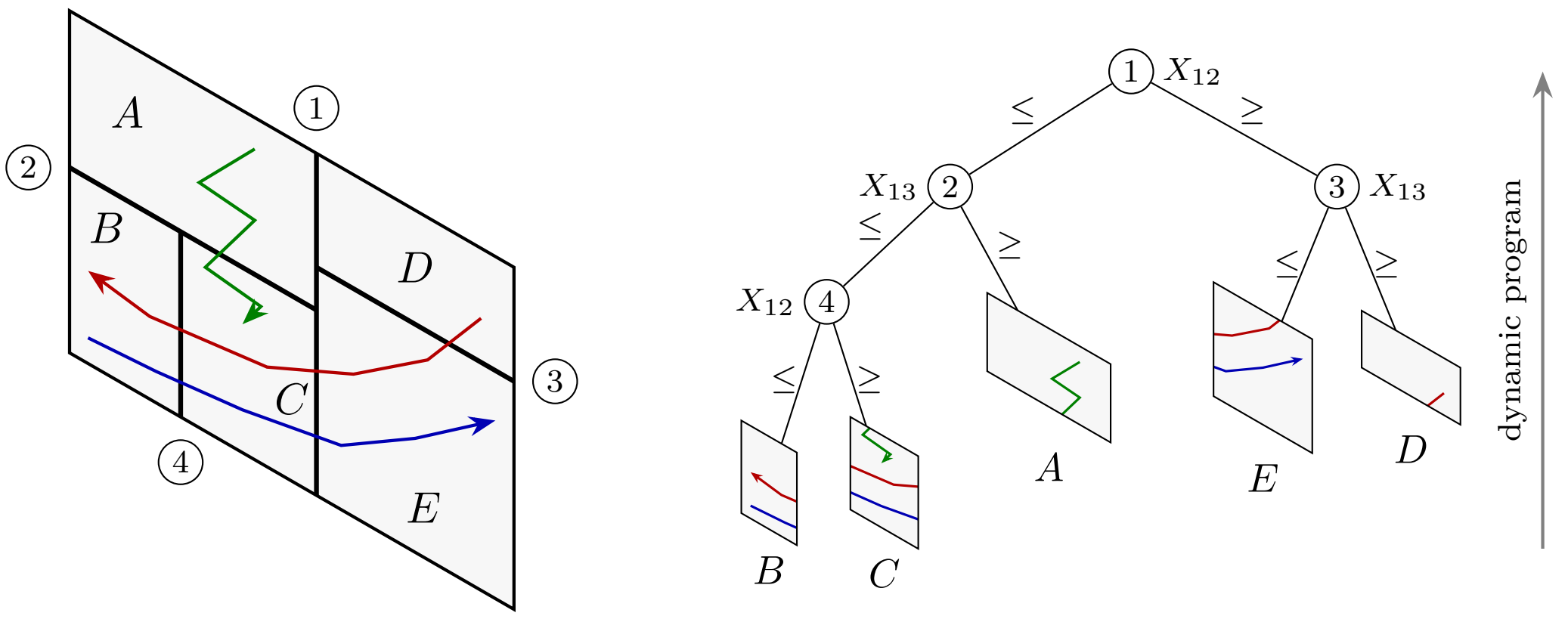
\begin{figure}[!tb]
  \centering
  \begin{tikzpicture}[scale=0.72,>=Stealth]
    \def\pone{(-2.475,-0.338) -- (-1.650,-0.745) -- (-0.600,-1.212) -- (0.600,-1.645)
      -- (1.500,-1.559) -- (2.475,-1.342)}
    \def\ptwo{(2.300,-0.100) -- (1.650,-0.606) -- (0.750,-0.779) -- (-0.300,-0.693)
      -- (-1.725,-0.078) -- (-2.475,0.476)}
    \def\pthree{(-0.450,1.957) -- (-1.125,1.550) -- (-0.450,1.091) -- (-1.050,0.520)
      -- (-0.375,0.043) -- (-0.600,-0.173)}
    \begin{scope}
      \def\rootcell{(-2.700,-0.520) -- (2.700,-3.637) -- (2.700,0.520) -- (-2.700,3.637) -- cycle}
      \fill[gray!6] \rootcell;
      \draw[very thick] (0.300,-2.252) -- (0.300,1.905);
      \draw[very thick] (-2.700,1.732) -- (0.300,0.000);
      \draw[very thick] (0.300,0.520) -- (2.700,-0.866);
      \draw[very thick] (-1.350,-1.299) -- (-1.350,0.953);
      \draw[thick] \rootcell;
      \draw[thick,blue!70!black,->] \pone;
      \draw[thick,red!70!black,->] \ptwo;
      \draw[thick,green!50!black,->] \pthree;
      \tikzset{cutlabel/.style={circle,draw,thin,inner sep=0.6pt,minimum size=11pt,
        font=\scriptsize}}
      \node[cutlabel] at (0.300,2.455) {$1$};
      \node[cutlabel] at (-3.200,1.732) {$2$};
      \node[cutlabel] at (3.200,-0.866) {$3$};
      \node[cutlabel] at (-1.350,-1.849) {$4$};
      \node at (-2.000,2.400) {$A$};
      \node at (-2.250,1.000) {$B$};
      \node at (0.000,-1.080) {$C$};
      \node at (1.500,0.520) {$D$};
      \node at (1.600,-2.400) {$E$};
      \node at (0,-4.1) {(a)};
    \end{scope}
    \begin{scope}[xshift=10.2cm,yshift=2.9cm]
      \tikzset{inner/.style={circle,draw,thin,inner sep=0.6pt,minimum size=11pt,font=\scriptsize},
        side/.style={font=\scriptsize,inner sep=1pt}}
      \node[inner] (n1) at (0,0) {$1$};
      \node[inner] (n2) at (-2.2,-1.4) {$2$};
      \node[inner] (n3) at (2.5,-1.4) {$3$};
      \node[inner] (n4) at (-3.7,-2.8) {$4$};
      \coordinate (A) at (-1.0,-3.6);
      \coordinate (B) at (-4.4,-5.0);
      \coordinate (C) at (-3.0,-5.0);
      \coordinate (D) at (3.4,-3.6);
      \coordinate (E) at (1.6,-3.6);
      \node[side,right=1pt] at (n1.east) {$\X{12}$};
      \node[side,left=1pt] at (n2.west) {$\X{13}$};
      \node[side,right=1pt] at (n3.east) {$\X{13}$};
      \node[side,left=1pt] at (n4.west) {$\X{12}$};
      \draw (n1) -- node[side,above left] {$\le$} (n2);
      \draw (n1) -- node[side,above right] {$\ge$} (n3);
      \draw (n2) -- node[side,above left] {$\le$} (n4);
      \draw (n2) -- node[side,above right,pos=0.35] {$\ge$} (A);
      \draw (n4) -- node[side,left,pos=0.35] {$\le$} (B);
      \draw (n4) -- node[side,right,pos=0.35] {$\ge$} (C);
      \draw (n3) -- node[side,left,pos=0.35] {$\le$} (E);
      \draw (n3) -- node[side,right,pos=0.35] {$\ge$} (D);
      \begin{scope}[shift={(-1.0,-3.6)},scale=0.5,shift={(1.2000,-1.8185)}]
        \fill[gray!6] (-2.700,1.732) -- (0.300,0.000) -- (0.300,1.905) -- (-2.700,3.637) -- cycle;
        \begin{scope}
          \clip (-2.700,1.732) -- (0.300,0.000) -- (0.300,1.905) -- (-2.700,3.637) -- cycle;
          \draw[semithick,blue!70!black,-{Stealth[length=3pt]}] \pone;
          \draw[semithick,red!70!black,-{Stealth[length=3pt]}] \ptwo;
          \draw[semithick,green!50!black,-{Stealth[length=3pt]}] \pthree;
        \end{scope}
        \draw[thin] (-2.700,1.732) -- (0.300,0.000) -- (0.300,1.905) -- (-2.700,3.637) -- cycle;
      \end{scope}
      \node[font=\small] at (-1.0,-4.81) {$A$};
      \begin{scope}[shift={(-4.4,-5.0)},scale=0.5,shift={(2.0250,-0.2165)}]
        \fill[gray!6] (-2.700,-0.520) -- (-1.350,-1.299) -- (-1.350,0.953) -- (-2.700,1.732) -- cycle;
        \begin{scope}
          \clip (-2.700,-0.520) -- (-1.350,-1.299) -- (-1.350,0.953) -- (-2.700,1.732) -- cycle;
          \draw[semithick,blue!70!black,-{Stealth[length=3pt]}] \pone;
          \draw[semithick,red!70!black,-{Stealth[length=3pt]}] \ptwo;
          \draw[semithick,green!50!black,-{Stealth[length=3pt]}] \pthree;
        \end{scope}
        \draw[thin] (-2.700,-0.520) -- (-1.350,-1.299) -- (-1.350,0.953) -- (-2.700,1.732) -- cycle;
      \end{scope}
      \node[font=\small] at (-4.4,-6.06) {$B$};
      \begin{scope}[shift={(-3.0,-5.0)},scale=0.5,shift={(0.5250,0.6495)}]
        \fill[gray!6] (-1.350,-1.299) -- (0.300,-2.252) -- (0.300,0.000) -- (-1.350,0.953) -- cycle;
        \begin{scope}
          \clip (-1.350,-1.299) -- (0.300,-2.252) -- (0.300,0.000) -- (-1.350,0.953) -- cycle;
          \draw[semithick,blue!70!black,-{Stealth[length=3pt]}] \pone;
          \draw[semithick,red!70!black,-{Stealth[length=3pt]}] \ptwo;
          \draw[semithick,green!50!black,-{Stealth[length=3pt]}] \pthree;
        \end{scope}
        \draw[thin] (-1.350,-1.299) -- (0.300,-2.252) -- (0.300,0.000) -- (-1.350,0.953) -- cycle;
      \end{scope}
      \node[font=\small] at (-3.0,-6.10) {$C$};
      \begin{scope}[shift={(3.4,-3.6)},scale=0.5,shift={(-1.5000,-0.5198)}]
        \fill[gray!6] (0.300,0.520) -- (2.700,-0.866) -- (2.700,0.520) -- (0.300,1.905) -- cycle;
        \begin{scope}
          \clip (0.300,0.520) -- (2.700,-0.866) -- (2.700,0.520) -- (0.300,1.905) -- cycle;
          \draw[semithick,blue!70!black,-{Stealth[length=3pt]}] \pone;
          \draw[semithick,red!70!black,-{Stealth[length=3pt]}] \ptwo;
          \draw[semithick,green!50!black,-{Stealth[length=3pt]}] \pthree;
        \end{scope}
        \draw[thin] (0.300,0.520) -- (2.700,-0.866) -- (2.700,0.520) -- (0.300,1.905) -- cycle;
      \end{scope}
      \node[font=\small] at (3.4,-4.59) {$D$};
      \begin{scope}[shift={(1.6,-3.6)},scale=0.5,shift={(-1.5000,1.5587)}]
        \fill[gray!6] (0.300,-2.252) -- (2.700,-3.637) -- (2.700,-0.866) -- (0.300,0.520) -- cycle;
        \begin{scope}
          \clip (0.300,-2.252) -- (2.700,-3.637) -- (2.700,-0.866) -- (0.300,0.520) -- cycle;
          \draw[semithick,blue!70!black,-{Stealth[length=3pt]}] \pone;
          \draw[semithick,red!70!black,-{Stealth[length=3pt]}] \ptwo;
          \draw[semithick,green!50!black,-{Stealth[length=3pt]}] \pthree;
        \end{scope}
        \draw[thin] (0.300,-2.252) -- (2.700,-3.637) -- (2.700,-0.866) -- (0.300,0.520) -- cycle;
      \end{scope}
      \node[font=\small] at (1.6,-4.94) {$E$};
      \draw[->,thick,gray] (5.0,-5.8) -- node[sloped,above,font=\scriptsize,text=black]
        {dynamic program} (5.0,0);
      \node at (0,-7.0) {(b)};
    \end{scope}
  \end{tikzpicture}
  \caption{The cut hierarchy for $k=3$.  The potentials place every vertex at a point of a
    plane, and the region of these points, the root cell, is cut recursively.
    (a) Each cut, numbered in order, is a line in one of two fixed directions (a level of $\X{12}$
    or of $\X{13}$) and goes across the whole cell being cut, so all cells are parallelograms
    (boxes in the coordinates $\X{12}$ and $\X{13}$).  (b) The cuts form a binary tree, as in a
    kd-tree: every inner node is a cut (marked with the coordinate it fixes), its children are the
    cells on its two sides, and the leaves are the cells $A,\dots,E$,
    each drawn with the pieces of the paths inside it.  The colored curves are three
    disjoint paths; each cut is crossed by them only a few times.  The recursion goes on
    until every cell contains only one point.  The dynamic program works through the tree from the
    leaves to the root.}
  \label{fig:kdtree}
\end{figure}

For $k$-DSPP the arcs that
lie on some shortest $s_i$--$t_i$ path form an acyclic graph $D_i$, but different pairs may cross
the same region in opposite directions, so $D_1\cup\dots\cup D_k$ has cycles and there is no
common order to sweep along.  Here we have to create the setting in which a dynamic program
works, and we replace the single order by a geometric one.  Potentials place every vertex at a
point of a $(k-1)$-dimensional space in which each path moves only forward in its own set of
directions.  We cut this space recursively into boxes, as in a kd-tree (see~\cref{fig:kdtree}), and a dynamic program over the tree
of boxes solves the problem; each cut is crossed by the paths only a few times, which gives the
small interface.  Inside a single point all pairs move in the same direction, which gives back
the common order that pebbling needs.

\paragraph{From shortest paths to monotone curves.}  A path from $s_i$ to $t_i$ is shortest if
and only if all its arcs lie in $D_i$, so a solution is a $k$-tuple of disjoint paths, the $i$-th
in $D_i$.  We give each pair $i$ a potential $F_i(v)$, essentially
$\dist(s_i,v)+(d_i-\dist(v,t_i))$ with $d_i=\dist(s_i,t_i)$, and map every vertex $v$ to the point
$\pt(v)=(F_1(v),\dots,F_k(v))\in\R^k$.  Since all pairs measure distance with the same length
function, one inequality holds for all of them at once \Wzero: along any arc $(u,v)$ every
potential rises by at most $2\ell(u,v)$, and $F_i$ rises by exactly $2\ell(u,v)$ if the arc lies
in $D_i$.  
Write $\X{ij}=F_i-F_j$.  Along an arc of $D_i$ therefore
\begin{itemize}
  \item the $k-1$ coordinates $\X{ij}$ ($j\neq i$) do not decrease (the \emph{monotone}
    coordinates of path $i$);
  \item every other coordinate $\X{hj}=\X{ij}-\X{ih}$ ($h,j\neq i$) may go up or down, but it
    moves by at most the sum of the moves of the two monotone coordinates $\X{ih}$ and $\X{ij}$.
\end{itemize}

\begin{figure}[!tb]
  \centering
  \begin{tikzpicture}[scale=0.62,>=Stealth]
    \begin{scope}
      \fill[blue!12] (0,0) -- (-30:2.4) arc (-30:90:2.4) -- cycle;
      \fill[red!12] (0,0) -- (90:2.4) arc (90:210:2.4) -- cycle;
      \fill[green!14] (0,0) -- (210:2.4) arc (210:330:2.4) -- cycle;
      \draw[gray] (-30:2.6) -- (150:2.6) (90:2.6) -- (270:2.6) (210:2.6) -- (30:2.6);
      \draw[->,thick,blue!70!black] (0,0) -- (30:2.2);
      \draw[->,thick,red!70!black] (0,0) -- (150:2.2);
      \draw[->,thick,green!50!black] (0,0) -- (270:2.2);
      \node[blue!70!black] at (30:3.0) {$1$};
      \node[red!70!black] at (150:3.0) {$2$};
      \node[green!50!black] at (270:3.0) {$3$};
      \draw[->] (-2.9,-2.9) -- (-1.9,-2.9) node[right] {\small $\X{12}$};
      \node at (0,-3.9) {(a)};
    \end{scope}
    \begin{scope}[xshift=9.5cm]
      \draw[thick,fill=gray!6] (-2.700,-0.520) -- (2.700,-3.637) -- (2.700,0.520)
        -- (-2.700,3.637) -- cycle;
      \draw[dashed,gray!70!black] (-1.200,-1.386) -- (-1.200,2.771);
      \draw[very thick] (0.900,-2.598) -- (0.900,1.559);
      \node[anchor=south west,inner sep=1pt] at (1.050,1.749) {\small $\X{12}=\lambda$};
      \draw[thick,blue!70!black,->] (-2.400,-0.346) -- (-1.500,-0.658) -- (-0.450,-1.091)
        -- (0.375,-0.944) -- (1.500,-1.299) -- (2.475,-1.516);
      \draw[thick,red!70!black,->] (2.475,-0.823) -- (1.500,-0.087) -- (0.450,0.779)
        -- (-0.600,1.992) -- (-1.650,2.650) -- (-2.475,2.815);
      \draw[thick,green!50!black,->] (-0.900,1.819) -- (-1.500,1.438) -- (-0.900,1.057)
        -- (-1.500,0.675) -- (-0.900,0.294) -- (-1.500,-0.087) -- (-0.900,-0.468)
        -- (1.650,-2.200);
      \foreach \p in {(0.900,-1.110),(0.900,0.408),(0.900,-1.690)} \fill \p circle (2.6pt);
      \node[blue!70!black] at (3.035,-1.516) {$1$};
      \node[red!70!black] at (-3.035,2.815) {$2$};
      \node[green!50!black] at (2.050,-2.350) {$3$};
      \node at (0,-3.9) {(b)};
    \end{scope}
    \begin{scope}[xshift=18.5cm]
      \def\rootcell{(-2.700,-0.520) -- (2.700,-3.637) -- (2.700,0.520) -- (-2.700,3.637) -- cycle}
      \def\cell{(-1.350,-1.299) -- (0.300,-2.252) -- (0.300,0.000) -- (-1.350,0.953) -- cycle}
      \def\pone{(-2.475,-0.338) -- (-1.650,-0.745) -- (-0.600,-1.212) -- (0.600,-1.645)
        -- (1.500,-1.559) -- (2.475,-1.342)}
      \def\ptwo{(2.300,-0.100) -- (1.650,-0.606) -- (0.750,-0.779) -- (-0.300,-0.693)
        -- (-1.725,-0.078) -- (-2.475,0.476)}
      \def\pthree{(-0.450,1.957) -- (-1.125,1.550) -- (-0.450,1.091) -- (-1.050,0.520)
        -- (-0.375,0.043) -- (-0.600,-0.173)}
      \fill[gray!6] \rootcell;
      \fill[yellow!35] \cell;
      \draw[thick,blue!70!black,opacity=0.3,->] \pone;
      \draw[thick,red!70!black,opacity=0.3,->] \ptwo;
      \draw[thick,green!50!black,opacity=0.3] \pthree;
      \begin{scope}
        \clip \cell;
        \draw[thick,blue!70!black] \pone;
        \draw[thick,red!70!black] \ptwo;
        \draw[thick,green!50!black] \pthree;
      \end{scope}
      \draw[very thick] (0.300,-2.252) -- (0.300,1.905);
      \draw[very thick] (-2.700,1.732) -- (0.300,0.000);
      \draw[very thick] (0.300,0.520) -- (2.700,-0.866);
      \draw[very thick] (-1.350,-1.299) -- (-1.350,0.953);
      \draw[thick] \rootcell;
      \tikzset{cutlabel/.style={circle,draw,thin,inner sep=0.6pt,minimum size=11pt,
        font=\scriptsize}}
      \node[cutlabel] at (0.300,2.455) {$1$};
      \node[cutlabel] at (-3.200,1.732) {$2$};
      \node[cutlabel] at (3.200,-0.866) {$3$};
      \node[cutlabel] at (-1.350,-1.849) {$4$};
      \foreach \p in {(-1.350,-0.878),(0.300,-0.742),(-0.880,0.681)} \fill \p circle (2.6pt);
      \draw[green!50!black,fill=white,thick] (-0.600,-0.173) circle (2.6pt);
      \node at (0,-3.9) {(c)};
    \end{scope}
  \end{tikzpicture}
  \caption{The case $k=3$, drawn in the plane $\R^3/\R\one$ with $\X{12}$ horizontal.
    (a) Along an arc of $D_i$ the point moves inside the shaded cone of path $i$: path $1$
    increases $\X{12}$ and $\X{13}$, path $2$ increases $\X{21}$ and $\X{23}$, path $3$
    increases $\X{31}$ and $\X{32}$.  The coordinate $\X{12}$ is free for path $3$.
    (b) A cell, bounded by two levels of $\X{12}$ and two levels of $\X{13}$,
    whose widest coordinate is $\X{12}$.  Path $3$ zigzags in its free coordinate, so the dashed
    level is crossed eight times (six times by path $3$); but each zigzag costs path $3$ travel in
    its monotone coordinates, which the cell bounds.  On average over the levels at most
    $3k-4=5$ crossings occur, and the cut is made at a good level (solid, three crossings).
    (c) The recursion of~\cref{fig:kdtree}, with paths colored as in (b): each cut, numbered in order, is a level of
    $\X{12}$ or $\X{13}$ and goes across the whole cell being cut, so the cells stay
    boxes, each with at most $2(k-1)=4$ walls.  The shaded cell, $C$ in~\cref{fig:kdtree}, is bounded by walls from
    cuts $1$, $2$ and $4$ and by the root cell.  Inside it each piece of a curve starts on a wall
    (dots) or at the beginning of its path; path $3$ ends inside the cell (circle).}
  \label{fig:sketch}
\end{figure}
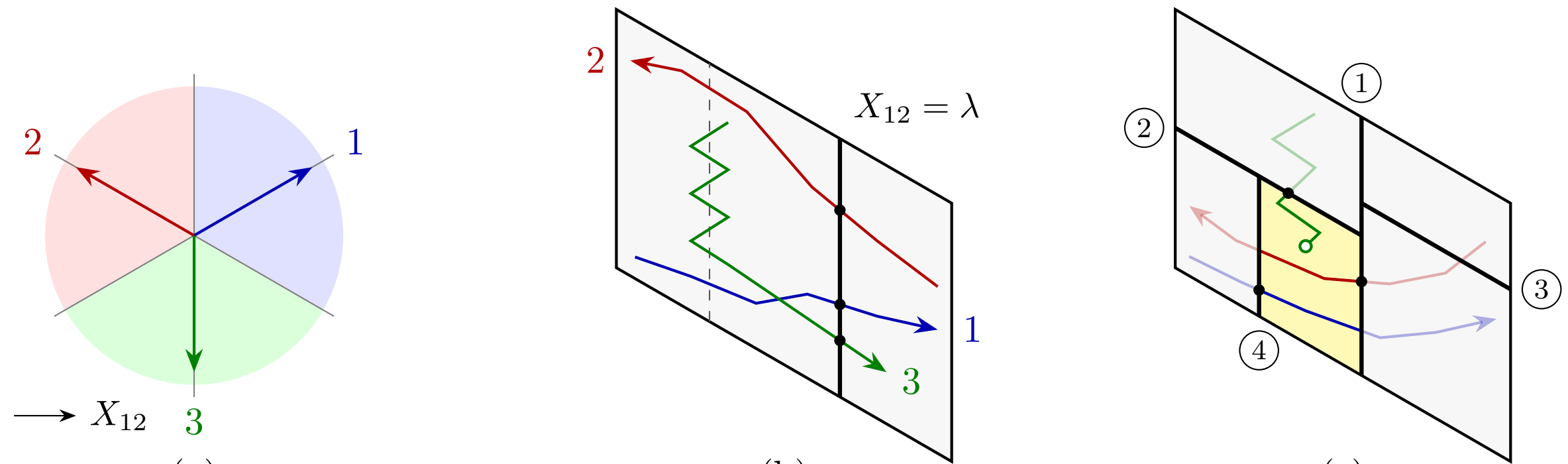

Adding the same constant to all potentials changes no $\X{ij}$.  The cuts that split the problem
into parts are hyperplanes $\X{ij}=\lambda$, made precise below; they use only the $\X{ij}$, since
these are the quantities that every path controls, by the two items above.  The common
shift is a clock on which all pairs agree; it causes no cycles, and it is used only inside a
single point.  So the picture lives in $\R^k/\R\one$, which has dimension $k-1$.
For $k=3$ it is a plane with three coordinate directions at $120^\circ$, and each path moves
inside a fixed $120^\circ$ cone of directions; see~\cref{fig:sketch}(a).  Drawing every path of a
solution as a polygonal curve through the points of its vertices, we obtain $k$ curves, the
$i$-th monotone in its $k-1$ own coordinates.

\paragraph{Cells and good cuts.}  We cut only along the $k-1$ coordinates $\X{1j}=F_1-F_j$,
$j=2,\dots,k$; they determine all the others, since $\X{ij}=\X{1j}-\X{1i}$.  A \emph{cell} is a
box $\{x: a_j\le \X{1j}(x)\le b_j\text{ for }j=2,\dots,k\}$ in these coordinates, with real bounds
$a_j\le b_j$; for $k=3$ it is bounded by two levels of $\X{12}$ and two levels of $\X{13}$.  The
bounds of the root cell, where the cutting starts, are the least and the largest value of $\X{1j}$ over all vertices, and
every cut replaces one bound by its level.

Let $C$ be a cell.  Choose a coordinate $\X{1j}$ for which the width $E_j=b_j-a_j$
is largest, and write $E=E_j$.  How often do the curves cross the hyperplane $\X{1j}=\lambda$
inside $C$, on average over $\lambda\in[a_j,b_j]$?  A straight segment of a curve inside $C$
whose $\X{1j}$ values run from $a$ to $b$ crosses each level strictly between $a$ and $b$ once.
It therefore contributes $|b-a|$ to the integral of the crossing count.  Summing over the
segments gives the total variation of $\X{1j}$ along the curves inside $C$.

Even if a path leaves and re-enters $C$, the variation in a monotone coordinate across all its
visits is bounded by the absolute difference between its values at the first and last points
in $C$.  Path $1$ is nondecreasing in $\X{1j}$, and path $j$ is nonincreasing in that same
coordinate, so each contributes at most $E$.

For any other path $i$, every movement
in $\X{1j}=\X{ij}-\X{i1}$ is paid for by movement in the monotone coordinates $\X{i1}$ and
$\X{ij}$ of path~$i$.  Over $C$ the range of $\X{i1}=-\X{1i}$ is $E_i\le E$, and the range of
$\X{ij}=\X{1j}-\X{1i}$ is at most $E_j+E_i\le 2E$, since the range of a difference is at most the
sum of the ranges; so path $i$ contributes at most $3E$.  The total is at most
$2E+(k-2)\cdot 3E=(3k-4)E$, so the average number of crossings is at most $3k-4$.
A level with more than $3k-4$ crossings has at least $3k-3$, since the count is an
integer.  Such levels can therefore occupy at most a fraction $(3k-4)/(3k-3)$ of the interval.
This leaves a level $\lambda$ with at most $3k-4$ crossings even after we trim a small fixed
fraction from both ends and exclude the finitely many levels attained by vertices
(\cref{cor:goodlevel}).

A path that zigzags across one hyperplane many times does not
contradict this: its zigzags cost it travel in its monotone coordinates, which is bounded by the
cell, so they cover only a small part of the range of $\X{1j}$; see~\cref{fig:sketch}(b).  The
choice of the widest coordinate is essential: for a narrow coordinate $\X{1j}$ the same argument
gives the ratio $(\text{range of }\X{i1}+\text{range of }\X{ij})/(\text{range of }\X{1j})$, which
is unbounded.

\paragraph{The cut hierarchy.}  Cutting a cell at a good level gives two smaller cells, and each
of them is cut again in the same way.  The cells obtained in this way form a binary tree: the root
cell is the root of the tree, every cut is an inner node, and its two children are the cells on
the two sides of the cut; see~\cref{fig:kdtree}.

This is much like a kd-tree~\cite{Bentley75}, which stores a set of points by splitting space
again and again with a hyperplane on which one coordinate is constant, in the variant that splits along a widest
coordinate~\cite{FBF77}.  There are two differences.  A kd-tree cuts at the median of the points, so that half of them go
to each side.  We cut instead at a level that the paths cross only a few times, since these
crossings are what the dynamic program will have to guess.  And since a kd-tree halves the points
at every cut, its depth is about $\log n$; our tree does not count points, and its depth is
governed by the lengths instead, as follows.

The good level lies in the middle part of the range of the widest coordinate, at distance at
least a fixed fraction of the range from both ends (\cref{cor:goodlevel}).  So each side of a cut
keeps at most a fixed fraction of the width that was cut, and since we always cut a widest
coordinate, the widths of the cells shrink geometrically along every branch of the tree.  The
coordinates $\X{1j}$ take only finitely many values on the vertices, so once a cell is narrower
than the smallest gap between these values, all its vertices have the same coordinates $\X{1j}$,
that is, they sit at the same point of the picture.  Such vertices form one \emph{point class}: a
set of vertices whose points differ by multiples of $(1,\dots,1)$, a shift that the picture does
not see.  There the recursion stops.  The number of cuts needed depends on the ratio between the
largest width and the smallest gap, which comes from the lengths and not from~$n$.

Every
cell is the root cell intersected with at most one lower and one upper bound per coordinate
$\X{1j}$, that is, with at most $2(k-1)$ \emph{walls}, each made by an earlier cut.  Each piece of
a curve inside the cell has two ends, and each end is an end of its path or a point where the
curve passes through a wall; that point lies in the (larger) cell in which the wall was cut, and
there the wall was crossed at most $3k-4$ times.  So every cell contains at most
$\bigl(2(k-1)(3k-4)+2k\bigr)/2=3k^2-6k+4$ pieces of the curves (\cref{lem:pieces}).

This is the structural theorem (\cref{thm:hierarchy}).  It
holds for every $k$-tuple of shortest paths, disjoint or not, and its bounds do not depend on
$n$; only the depth of the recursion depends on the lengths, and the algorithm never sees it.

\paragraph{The dynamic program.}  The algorithm follows the pattern of the approximation schemes
of Arora~\cite{Arora98} and Mitchell~\cite{Mitchell99} for the Euclidean TSP.  These are dynamic
programs over a recursive subdivision: a solution crosses the boundary of every region only a few
times, and the table guesses these crossings.  There the subdivision is fixed, and the
solution is modified to cross it rarely.  Here a solution cannot be modified, and its hierarchy
depends on it.

The algorithm does not know a solution, so it cannot build its hierarchy; it tries all
possibilities instead.  A table entry consists of a set $U$ of
vertices of the form $\{v:\alpha_j\le \X{1j}(\pt(v))\le\beta_j\text{ for }j=2,\dots,k\}$ with
bounds that are vertex values (at most $n^{2(k-1)}$ sets; the vertex set of every cell is of this
form), and a list of at most $3k^2-6k+4$ \emph{demands} $(i,a,b)$ with $a,b\in U$.  To evaluate
an entry, the table tries every split of $U$ by one coordinate $\X{1j}$ into two parts.  For each
split it tries every choice of at most $3k-4$ arcs between the two parts, each assigned to a
demand.  The chosen arcs cut every demand into sub-demands on the two sides, and the entry is true
if both sides are.

A true entry certifies that its demands can be routed by pairwise disjoint paths
inside $U$, with the path for $(i,a,b)$ in $D_i$.  At a split, the crossing arcs join the
disjoint subpaths on the two sides.  In particular, a true root entry gives disjoint shortest
paths.  If a solution exists, the structural theorem supplies a hierarchy whose splits and
crossing arcs make every entry it induces true.

At the bottom, $U$ is
a single point class.  Inside a point class every usable arc raises all potentials by the same
positive amount \Wtwo, so the union of the arc sets is acyclic, every shortest path visits the
class in one contiguous piece, and the pebbling game of~\cite{FHW80} decides the entry.  Counting
entries and choices gives $\smash[t]{n^{O(k^2)}}$.

\paragraph{Why not a sweep.}  A natural first attempt is to process the space one level of a
fixed coordinate at a time.  This fails because a fixed hyperplane can be crossed by a solution
arbitrarily often by the free paths of that coordinate.  The structural theorem works because it \emph{chooses} the coordinate (a
widest one) and the level (a good one in the middle), and because cells cut by the hyperplanes
$\X{1j}=\lambda$ remain cells with at most $2(k-1)$ walls.

\section{Preliminaries}\label{sec:prelim}

\subsection{Instances and shortest-path DAGs}

Throughout, $k\ge 2$, $G=(V,A)$ is a directed graph with $n=|V|$ vertices and $m=|A|$ arcs,
$\ell\colon A\to\R_{>0}$ is a length function, and $(s_1,t_1),\dots,(s_k,t_k)$ are terminal pairs
with $2k$ distinct terminals.  A \emph{path} is a sequence $P=(v_0,\dots,v_r)$, $r\ge 0$, of
distinct vertices with $(v_{q-1},v_q)\in A$ for $q=1,\dots,r$; these $r$ pairs are its arcs, and
its length $\ell(P)$ is the sum of their lengths.  It is a \emph{$u$--$v$ path} if $v_0=u$ and
$v_r=v$; for $r=0$ it consists of a single vertex.  Paths in undirected graphs are defined in the
same way, with edges in place of arcs.

We write $\dist(u,v)\in\R_{\ge0}\cup\{\infty\}$ for the
distance from $u$ to $v$, the least length of a $u$--$v$ path, and
$d_i=\dist(s_i,t_i)$.  If some $d_i$ is infinite the answer is trivially no, so we assume that all
$d_i$ are finite.  A \emph{solution} is a $k$-tuple of pairwise vertex-disjoint paths
$P_1,\dots,P_k$ such that $P_i$ is a shortest $s_i$--$t_i$ path.  This is the vertex-disjoint
version of $k$-DSPP, the one in~\cref{thm:main}, which~\cref{sec:prelim,sec:hierarchy,sec:algorithm}
prove; \cref{sec:remarks} reduces the arc- and edge-disjoint versions to it.

Indices $h,i,j$ range
over $\{1,\dots,k\}$; each names a path, and the $j$-th coordinate of a point is the potential
$F_j$ of path $j$.  When a statement fixes one path we usually call it $i$, and $h,j$ denote the
others.

\begin{definition}\label{def:Di}
  For each $i$ let $D_i$ be the set of arcs $(u,v)\in A$ with
  \[
    \dist(s_i,u)+\ell(u,v)+\dist(v,t_i)=d_i.
  \]
  A \emph{$D_i$-path} is a path all of whose arcs
  lie in $D_i$.
\end{definition}

\begin{lemma}\label{lem:Di}
  \leavevmode
  \begin{enumerate}[label=(\alph*)]
    \item Along every arc $(u,v)\in D_i$ we have $\dist(s_i,v)=\dist(s_i,u)+\ell(u,v)\le d_i$ and
      $\dist(u,t_i)=\ell(u,v)+\dist(v,t_i)\le d_i$.
    \item A path from $s_i$ to $t_i$ is a shortest path if and only if it is a $D_i$-path.
    \item $D_i$ is acyclic.
  \end{enumerate}
\end{lemma}

\begin{proof}
  (a) By the triangle inequality $\dist(s_i,v)\le\dist(s_i,u)+\ell(u,v)$ and
  $d_i\le\dist(s_i,v)+\dist(v,t_i)$.  Together with the defining equation of $D_i$ this gives
  \[
    \dist(s_i,u)+\ell(u,v)+\dist(v,t_i)=d_i\le\dist(s_i,v)+\dist(v,t_i),
  \]
  so $\dist(s_i,v)\ge\dist(s_i,u)+\ell(u,v)$, hence equality; and $\dist(s_i,v)\le d_i$ because
  $\dist(v,t_i)\ge 0$.  The second statement is symmetric.
  (b) If all arcs of an $s_i$--$t_i$ path $P$ lie in $D_i$, then by (a) the length of each arc
  equals the increase of $\dist(s_i,\cdot)$ along it, and the sum telescopes to
  $\ell(P)=\dist(s_i,t_i)-\dist(s_i,s_i)=d_i$.  Conversely, if $P$ is shortest and $(u,v)$ is an arc of
  $P$, then the parts of $P$ before $u$ and after $v$ are shortest paths, so
  $\dist(s_i,u)+\ell(u,v)+\dist(v,t_i)=\ell(P)=d_i$.
  (c) Along a directed cycle of $D_i$ the increases of $\dist(s_i,\cdot)$ sum to $0$, but by (a) they
  equal the arc lengths, which are positive.
\end{proof}

So a solution is the same as a $k$-tuple of pairwise vertex-disjoint paths, the $i$-th an
$s_i$--$t_i$ path in $D_i$.  The union $D_1\cup\dots\cup D_k$ may contain directed cycles, and an
arc may belong to several $D_i$.  We may also assume that $G$ has no parallel arcs: of several
arcs from $u$ to $v$ keep one of minimum length; this changes no distance, and a solution using a
deleted arc can use the kept one instead (it then has the same length, so the path stays
shortest).  Hence $m\le n^2$.

\subsection{Potentials}

\begin{definition}[potentials]\label{def:potentials}
  For each $i$ and $v\in V$ let
  \[
    F_i(v)=\min\bigl(\dist(s_i,v),\,d_i\bigr)+\max\bigl(d_i-\dist(v,t_i),\,0\bigr),
  \]
  with $\min(\infty,d_i)=d_i$ and $\max(-\infty,0)=0$.
\end{definition}

\begin{lemma}[\textup{W0}]\label{lem:W0}
  For every arc $(u,v)\in A$ and every $j$,
  \[
    F_j(v)-F_j(u)\le 2\ell(u,v),\qquad\text{with equality if }(u,v)\in D_j .
  \]
\end{lemma}

\begin{proof}
  By the triangle inequality $\dist(s_j,v)\le\dist(s_j,u)+\ell(u,v)$, also when distances are
  infinite.  Taking the minimum with the constant $d_j$ on both sides keeps the inequality, and
  $\min(a+\ell,d)\le\min(a,d)+\ell$ for $\ell\ge0$; so $v\mapsto\min(\dist(s_j,v),d_j)$ increases by at
  most $\ell(u,v)$ along $(u,v)$.  Similarly $\dist(u,t_j)\le\ell(u,v)+\dist(v,t_j)$ gives
  $d_j-\dist(v,t_j)\le d_j-\dist(u,t_j)+\ell(u,v)$, and taking the maximum with $0$ keeps this; so
  the second term also increases by at most $\ell(u,v)$.  If $(u,v)\in D_j$, then by~\cref{lem:Di}(a) the four distances $\dist(s_j,u),\dist(s_j,v),\dist(u,t_j),\dist(v,t_j)$ are at
  most $d_j$, so the minimum and the maximum are attained by the distance terms at both ends, and
  each term increases by exactly $\ell(u,v)$.
\end{proof}

Property \Wzero{} is the only property of the potentials used in the rest of the paper.  It is
also the only place where it matters that all pairs measure distances with the same length
function: the same number $2\ell(u,v)$ bounds the increase of all $k$ potentials.

\subsection{Points, coordinates and monotonicity}

The \emph{point} of a vertex $v$ is $\pt(v)=(F_1(v),\dots,F_k(v))\in\R^k$.  For $i\neq j$ the
\emph{coordinate} $\X{ij}\colon\R^k\to\R$ is $\X{ij}(x)=x_i-x_j$; so $\X{ji}=-\X{ij}$ and
$\X{ij}=\X{hj}-\X{hi}$ for all $h$.  We write $\X{ij}(v)$ for $\X{ij}(\pt(v))$.  All coordinates are
invariant under adding a multiple of $\one=(1,\dots,1)$.  The $k-1$ coordinates
$\X{12},\dots,\X{1k}$ are the \emph{star coordinates}.  They determine all other coordinates:
$\X{i1}=-\X{1i}$, and $\X{ij}=\X{1j}-\X{1i}$ for $i,j\neq1$.  Two vertices $u,v$ are in the same
\emph{point class} if $\pt(u)-\pt(v)\in\R\one$, that is, if $\X{ij}(u)=\X{ij}(v)$ for all $i,j$; by
the identities above this holds if and only if $u$ and $v$ agree in all star coordinates.

\begin{lemma}\label{lem:W12}
  Let $(u,v)\in D_i$ and $\Delta\X{hj}=\X{hj}(v)-\X{hj}(u)$.
  \begin{enumerate}[label=\textup{(W\arabic*)},start=1]
    \item $\Delta\X{ij}\ge 0$ for every $j\ne i$, and
      $\lvert\Delta\X{hj}\rvert\le\Delta\X{ih}+\Delta\X{ij}$ for all $h,j\neq i$.
    \item If $u$ and $v$ are in the same point class, then every potential $F_j$ increases by
      exactly $2\ell(u,v)>0$ along $(u,v)$.
  \end{enumerate}
\end{lemma}

\begin{proof}
  Let $\delta=\pt(v)-\pt(u)$.  By \Wzero{} $\delta_i=2\ell(u,v)\ge\delta_j$ for all $j$, so
  $\Delta\X{ij}=\delta_i-\delta_j\ge0$.  For $h,j\neq i$,
  $\Delta\X{hj}=\delta_h-\delta_j=(\delta_i-\delta_j)-(\delta_i-\delta_h)$ is the difference of
  two non-negative numbers, so its absolute value is at most their sum.  If $u,v$ are in the same
  point class, $\delta=c\one$ for some $c$, and $c=\delta_i=2\ell(u,v)$.
\end{proof}

Thus along a $D_i$-path the $k-1$ coordinates $\X{ij}$, $j\neq i$, are non-decreasing (the
\emph{monotone coordinates} of path $i$), and every other coordinate $\X{hj}$, $h,j\ne i$, can go
up and down, but only by as much as the two monotone coordinates $\X{ih}$ and $\X{ij}$ travel.
Note that $\X{ij}$ is non-increasing along every $D_j$-path, since $\X{ij}=-\X{ji}$.  In
particular, along a $D_1$-path all star coordinates are non-decreasing; along a $D_i$-path with
$i\neq1$ the star coordinate $\X{1i}$ is non-increasing, and every other star coordinate $\X{1j}$
changes by at most as much as $\X{i1}$ and $\X{ij}$ travel.

\begin{lemma}\label{lem:classpiece}
  Let $Q$ be a point class and $R=(v_0,\dots,v_r)$ a $D_i$-path.  The indices $q$ with $v_q\in Q$
  form an interval of integers.
\end{lemma}

\begin{proof}
  Let $a<b<c$ with $v_a,v_c\in Q$.  By \Wone, $\X{ij}(v_a)\le\X{ij}(v_b)\le\X{ij}(v_c)=\X{ij}(v_a)$
  for every $j\ne i$, so $\X{ij}(v_b)=\X{ij}(v_a)$ for all $j$, and then
  $\X{hj}(v_b)=\X{ij}(v_b)-\X{ih}(v_b)=\X{hj}(v_a)$ for all $h,j$.  So $v_b\in Q$.
\end{proof}

\subsection{Cells}

\begin{definition}[cells]\label{def:cell}
  A \emph{cell} is a set
  $C=\{x\in\R^k: a_j\le\X{1j}(x)\le b_j\text{ for }j=2,\dots,k\}$ with real bounds $a_j\le b_j$.
  Its \emph{extent} in a coordinate $\X{ij}$ is $\ext_{ij}(C)=\max_C\X{ij}-\min_C\X{ij}$ (so
  $\ext_{ji}=\ext_{ij}$), and its \emph{width} $w(C)=\max_{2\le j\le k}\ext_{1j}(C)$ is the largest
  extent of a star coordinate.  We write $V(C)=\{v\in V:\pt(v)\in C\}$.
\end{definition}

\begin{lemma}[extent formula]\label{lem:extent}
  Let $C$ be the cell with bounds $a_j\le b_j$, and let $E_j=b_j-a_j$ for $2\le j\le k$.
  \begin{enumerate}[label=(\alph*)]
    \item The map $x\mapsto(\X{12}(x),\dots,\X{1k}(x))$ maps $C$ onto the box
      $B=[a_2,b_2]\times\dots\times[a_k,b_k]$.  In particular $C\neq\emptyset$, and $C$ is closed,
      convex and invariant under adding multiples of $\one$.
    \item $\ext_{1j}(C)=E_j$ for $2\le j\le k$, and $\ext_{ij}(C)=E_i+E_j$ for $2\le i<j\le k$.
      Hence $w(C)=\max_jE_j$, every extent of $C$ is at most $2w(C)$, and the bounds are
      determined by $C$: $a_j=\min_C\X{1j}$ and $b_j=\max_C\X{1j}$.
    \item Let $2\le j\le k$ and $\lambda\in\R$.  If $\lambda\ge a_j$, then
      $C\cap\{\X{1j}\le\lambda\}$ is the cell with the same bounds as $C$ except that $b_j$ is
      replaced by $\min(b_j,\lambda)$; otherwise it is empty.  If $\lambda\le b_j$, then
      $C\cap\{\X{1j}\ge\lambda\}$ is the cell with the same bounds as $C$ except that $a_j$ is
      replaced by $\max(a_j,\lambda)$; otherwise it is empty.
  \end{enumerate}
\end{lemma}

\begin{proof}
  (a) All coordinates are linear and invariant under adding multiples of $\one$, and $C$ is
  defined by non-strict inequalities on coordinates; so $C$ is closed, convex and invariant.  By
  definition the map sends $C$ into $B$.  Conversely, for $y=(y_2,\dots,y_k)\in B$ the point
  $x=(0,-y_2,\dots,-y_k)$ has $\X{1j}(x)=0-(-y_j)=y_j$ for all $j$, so $x\in C$ and $x$ is mapped
  to $y$.  As $a_j\le b_j$ for all $j$, $B$ and hence $C$ are non-empty.

  (b) By (a), $\X{1j}$ takes on $C$ exactly the values in $[a_j,b_j]$; this gives
  $\ext_{1j}(C)=E_j$ and the last statement.  Let $2\le i<j\le k$.  Since $\X{ij}=\X{1j}-\X{1i}$, by
  (a) the values of $\X{ij}$ on $C$ are exactly the numbers $y_j-y_i$ with $y\in B$.  As $y_i$ and
  $y_j$ range independently over $[a_i,b_i]$ and $[a_j,b_j]$, the largest of these numbers is
  $b_j-a_i$ and the smallest is $a_j-b_i$, so $\ext_{ij}(C)=(b_j-a_i)-(a_j-b_i)=E_i+E_j$.  With
  $\ext_{ji}=\ext_{ij}$ this covers all pairs, and $E_i+E_j\le2w(C)$, as $w(C)$ is the largest of
  $E_2,\dots,E_k$.

  (c) The set $C\cap\{\X{1j}\le\lambda\}$ is defined by the bounds of $C$ with $b_j$ replaced by
  $\min(b_j,\lambda)$.  If $\lambda\ge a_j$, these are the bounds of a cell, since
  $a_j\le\min(b_j,\lambda)$; if $\lambda<a_j$, no point satisfies $a_j\le\X{1j}(x)\le\lambda$.  The
  second statement is symmetric.
\end{proof}

So a cell is a box in the $k-1$ star coordinates, its extents in the star coordinates are the
side lengths of the box, and its width is the longest side.  Sets defined by bounds on all
coordinates $\X{ij}$ are known as \emph{alcoved polytopes} (of type A)~\cite{LamPostnikov07}, up
to the translations along $\one$; a cell is the special case in which only the star coordinates
are bounded.  Since a cell is invariant under adding multiples of $\one$, $V(C)$ is a union of
whole point classes.

\subsection{Curves and generic levels}

For a $D_i$-path $R=(v_0,\dots,v_r)$ let $\gamma_R\colon[0,r]\to\R^k$ be the polygonal curve with
$\gamma_R(q)=\pt(v_q)$ for integers $q$, linear on every interval $[q,q+1]$.  The integers
$q\in[0,r]$ are the \emph{vertex parameters} of the curve.  On every linear piece of $\gamma_R$
the coordinates are linear functions, so \Wone{} gives, for $t\le t'$:
\begin{enumerate}[label=\textup{(M\arabic*)}]
  \item $\X{ij}(\gamma_R(t))\le\X{ij}(\gamma_R(t'))$ for every $j\neq i$ (for all $t\le t'$ in
    $[0,r]$, not only in one linear piece);
  \item if $t,t'$ lie in one linear piece and $h,j\neq i$, then
    \begin{multline*}
      \lvert\X{hj}(\gamma_R(t'))-\X{hj}(\gamma_R(t))\rvert\\
      \le\bigl[\X{ih}(\gamma_R(t'))-\X{ih}(\gamma_R(t))\bigr]
      +\bigl[\X{ij}(\gamma_R(t'))-\X{ij}(\gamma_R(t))\bigr].
    \end{multline*}
\end{enumerate}

A value $\lambda$ is \emph{generic} for the coordinate $\X{hj}$ if no vertex $v$ has
$\X{hj}(v)=\lambda$; all but at most $n$ values are generic.  If $\lambda$ is generic and
$\X{hj}(\gamma_R(t))=\lambda$, then $t$ is not a vertex parameter.  So $t$ is an interior point of
a linear piece on which $\X{hj}\circ\gamma_R$ is not constant (a constant piece takes the value of
its end vertices).  Hence the curve crosses the hyperplane $\{\X{hj}=\lambda\}$ transversally at
$t$, and it meets this hyperplane at most once in each linear piece.

\begin{definition}[crossing number]\label{def:N}
  Let $\mathcal R=(R_1,\dots,R_k)$ be a $k$-tuple where $R_i$ is a $D_i$-path, with curves
  $\gamma_i=\gamma_{R_i}\colon[0,r_i]\to\R^k$.  For a cell $C$, a star coordinate $\X{1j}$ and a
  value $\lambda$ that is generic for $\X{1j}$ let
  \[
    N_C(\lambda)=\bigl\lvert\{(i,t): 1\le i\le k,\ t\in[0,r_i],\ \gamma_i(t)\in C,\
    \X{1j}(\gamma_i(t))=\lambda\}\bigr\rvert .
  \]
\end{definition}

The coordinate is always clear from the context.  By the last paragraph $N_C(\lambda)$ is finite
(at most the total number of arcs of the paths).  Note that $N_C$ counts points of the curves,
including points on segments whose end vertices lie outside $C$.

\section{The cut hierarchy}\label{sec:hierarchy}

In this section $\mathcal R=(R_1,\dots,R_k)$ is a fixed $k$-tuple where $R_i=(v^i_0,\dots,v^i_{r_i})$
is a $D_i$-path, with curves $\gamma_i=\gamma_{R_i}$.  The paths may have arbitrary end vertices
and need not be disjoint; in~\cref{sec:algorithm} we apply the results to a solution.  For a set
$U\subseteq V$, a \emph{piece} of $R_i$ in $U$ is a maximal set of consecutive vertices
$v^i_a,v^i_{a+1},\dots,v^i_b$ of $R_i$ that all lie in $U$, regarded as a subpath of $R_i$.  Let
\[
  \Pk=(k-1)(3k-4)+k=3k^2-6k+4 .
\]

\begin{definition}[hierarchy]\label{def:hierarchy}
  A \emph{hierarchy} is a finite rooted binary tree whose nodes are cells, such that every
  non-leaf cell $C$ has two children
  $C^-=C\cap\{\X{1j}\le\lambda\}$ and $C^+=C\cap\{\X{1j}\ge\lambda\}$ for some star coordinate
  $\X{1j}$, $2\le j\le k$, and some value $\lambda$ that is generic for $\X{1j}$ and satisfies
  $\min_C\X{1j}<\lambda<\max_C\X{1j}$.  We say that $C$ is \emph{cut} at $(1,j,\lambda)$.
\end{definition}

Since $\lambda$ is generic, no vertex point lies on the hyperplane $\{\X{1j}=\lambda\}$, so $V(C)$
is the disjoint union of $V(C^-)$ and $V(C^+)$.  By~\cref{lem:extent} the range of $\X{1j}$ on $C$
is $[a_j,b_j]$, where $a_j,b_j$ are the bounds of $C$ in $\X{1j}$, so $a_j<\lambda<b_j$, and $C^-$
and $C^+$ are the cells obtained from $C$ by replacing $b_j$, respectively $a_j$, by $\lambda$.  In
particular both children are non-empty as point sets ($V(C^-)$ or $V(C^+)$ may be empty).

\begin{theorem}[cut hierarchy]\label{thm:hierarchy}
  For every $k$-tuple $\mathcal R$ of $D_i$-paths as above there is a hierarchy such that
  \begin{enumerate}[label=\textup{(H\arabic*)}]
    \item the root cell contains $\pt(v)$ for every $v\in V$;
    \item\label{H:cut} if a cell $C$ is cut at $(1,j,\lambda)$, then $N_C(\lambda)\le 3k-4$ (with
      respect to the coordinate $\X{1j}$);
    \item\label{H:pieces} for every cell $C$ of the hierarchy, the sets $\gamma_i^{-1}(C)$,
      $i=1,\dots,k$, have at most $\Pk$ connected components in total; in particular the paths
      $R_1,\dots,R_k$ have at most $\Pk$ pieces in $V(C)$ in total;
    \item\label{H:leaf} a cell is a leaf if and only if $V(C)$ meets at most one point class.
  \end{enumerate}
\end{theorem}

\begin{corollary}\label{cor:crossarcs}
  In the hierarchy of~\cref{thm:hierarchy}, if $C$ is cut into $C^-$ and $C^+$, then at most
  $3k-4$ arcs of $R_1,\dots,R_k$ (counted with multiplicity over the paths) have one end in
  $V(C^-)$ and the other in $V(C^+)$.
\end{corollary}

\begin{proof}
  Let $C$ be cut at $(1,j,\lambda)$ and let $(v^i_q,v^i_{q+1})$ be an arc of $R_i$ with one end in
  $V(C^-)$ and the other in $V(C^+)$.  Both ends lie in $C$, so the segment $\gamma_i([q,q+1])$ lies
  in $C$ by convexity.  The affine function $\X{1j}\circ\gamma_i$ on $[q,q+1]$ has one end value
  below $\lambda$ and the other above, so it takes the value $\lambda$ at exactly one $t\in(q,q+1)$.
  Different arcs give different pairs $(i,t)$.  So the number of such arcs is at most
  $N_C(\lambda)\le 3k-4$ by~\ref{H:cut}.
\end{proof}

\subsection{The averaging lemma}

\begin{lemma}[averaging]\label{lem:averaging}
  Let $C$ be a cell with width $E=w(C)>0$, let $2\le j\le k$ be an index with $\ext_{1j}(C)=E$,
  and let $[m,M]$ be the range of $\X{1j}$ on $C$.  Then $N_C(\lambda)$ (with respect to $\X{1j}$)
  agrees, for all but finitely many $\lambda$, with a non-negative step function whose integral
  over $[m,M]$ is at most $(3k-4)E$.
\end{lemma}

\begin{proof}
  Fix $i$ with $r_i\ge1$ (a path with one vertex has a one-point curve, whose only value is a
  vertex value, so it contributes nothing to $N_C(\lambda)$ for generic $\lambda$).  For each
  integer $0\le q<r_i$ the map $\gamma_i$ is affine on $[q,q+1]$ and $C$ is closed and convex, so
  $\gamma_i^{-1}(C)\cap[q,q+1]$ is a closed interval or empty.  Let $\mathcal J_i$ be the family of
  these non-empty intervals; two of them share at most a vertex parameter.

  For $J=[t,t']\in\mathcal
  J_i$, write $f=\X{1j}\circ\gamma_i$ and $\Delta_J\X{1j}=f(t')-f(t)$, and similarly for other
  coordinates.  The function $f$ is affine on $J$.  For a generic $\lambda$ that is different from
  $f(t)$ and $f(t')$, the number of $\tau\in J$ with $f(\tau)=\lambda$ is $1$ if $\lambda$ lies
  strictly between $f(t)$ and $f(t')$ and $0$ otherwise (if $f$ is constant on $J$ and $t<t'$, its
  value is a vertex value and not generic).  No pair $(i,\tau)$ with $f(\tau)$ generic is counted in
  two intervals, because shared points are vertex parameters.  Hence, for all generic $\lambda$
  outside the finite set of end values $f(t),f(t')$,
  \[
    N_C(\lambda)=\sum_{i}\sum_{J\in\mathcal J_i}\mathbf 1\bigl[\lambda\text{ lies strictly between
    the two end values of }\X{1j}\circ\gamma_i\text{ on }J\bigr],
  \]
  a step function whose integral is $\sum_i\sum_{J\in\mathcal J_i}\lvert\Delta_J\X{1j}\rvert$.  We
  bound the inner sum for each $i$.

  First we show: if $g=\X{ih}\circ\gamma_i$ for some $h\ne i$ (a monotone coordinate of path $i$),
  then $\sum_{J\in\mathcal J_i}\Delta_Jg\le\ext_{ih}(C)$.  Indeed, list the intervals of $\mathcal
  J_i$ from left to right as $[t_1,t'_1],\dots,[t_s,t'_s]$, so $t'_q\le t_{q+1}$.  By \Mone{}, $g$ is
  non-decreasing on $[0,r_i]$, so $g(t'_q)\le g(t_{q+1})$ and
  \[
    \sum_{q=1}^{s}\bigl(g(t'_q)-g(t_q)\bigr)\le g(t'_s)-g(t_1)\le\ext_{ih}(C),
  \]
  because $\gamma_i(t_1)$ and $\gamma_i(t'_s)$ lie in $C$.

  \emph{Case $i=1$.}  Then $\X{1j}=\X{ij}$ is a monotone coordinate of path $i$, every
  $\Delta_J\X{1j}$ is non-negative, and the sum is at most $\ext_{1j}(C)=E$.

  \emph{Case $i=j$.}  Then $\X{1j}=-\X{i1}$, every $\Delta_J\X{1j}$ is non-positive, and
  $\sum_J\lvert\Delta_J\X{1j}\rvert=\sum_J\Delta_J\X{i1}$, which is at most
  $\ext_{i1}(C)=\ext_{1j}(C)=E$.

  \emph{Case $i\notin\{1,j\}$.}  By \Mtwo{} (with the two indices $1$ and $j$, both different from
  $i$), $\lvert\Delta_J\X{1j}\rvert\le\Delta_J\X{i1}+\Delta_J\X{ij}$, so the sum is at most
  $\ext_{i1}(C)+\ext_{ij}(C)$.  By~\cref{lem:extent}, $\ext_{i1}(C)=E_i$ and
  $\ext_{ij}(C)=E_i+E_j$, where $E_i=\ext_{1i}(C)\le E$ because $E$ is the width of $C$, and
  $E_j=\ext_{1j}(C)=E$.  So the sum is at most $3E$.

  Adding up, the integral is at most $E+E+(k-2)\cdot 3E=(3k-4)E$.
\end{proof}

The hypothesis that $\X{1j}$ is a widest star coordinate is used only in the last case, and it is
essential there: for a star coordinate of small extent, the curve of a path $i\notin\{1,j\}$ may
cross the slab $\{m\le\X{1j}\le M\}$ about $(\ext_{i1}(C)+\ext_{ij}(C))/\ext_{1j}(C)$ times.  In the
last case a path contributes up to $3E$, not $2E$, because the extent of $\X{ij}$, which is not a
star coordinate, can be as large as $2E$.

\begin{corollary}[good level]\label{cor:goodlevel}
  In the situation of~\cref{lem:averaging}, let $\theta=1/(6k)$.  There is a value $\lambda$,
  generic for $\X{1j}$, with $m+\theta E\le\lambda\le M-\theta E$ and $N_C(\lambda)\le 3k-4$.
\end{corollary}

\begin{proof}
  Let $\phi$ be the step function of~\cref{lem:averaging}.  Since $\phi\ge 0$, the set
  \[
    \{\lambda\in[m,M]:\phi(\lambda)\ge 3k-3\}
  \]
  has measure at most
  \[
    (3k-4)E/(3k-3)=\bigl(1-\frac{1}{3k-3}\bigr)E,
  \]
  which is less than $\bigl(1-\frac{1}{3k}\bigr)E=(1-2\theta)E$, the length of the interval
  $I=[m+\theta E,M-\theta E]$.  So $\phi\le 3k-4$ on a subset of $I$ of positive measure.  Only
  finitely many values are not generic or are exceptional in~\cref{lem:averaging}, so this subset
  contains a generic $\lambda$ with $N_C(\lambda)=\phi(\lambda)\le 3k-4$.
\end{proof}

\subsection{Construction and termination}

\begin{definition}[the hierarchy of $\mathcal R$]\label{def:construction}
  The root is the cell
  \[
    C_0=\bigl\{x\in\R^k:\ \min_{v\in V}\X{1j}(v)\le\X{1j}(x)\le\max_{v\in V}\X{1j}(v)\text{ for }
    j=2,\dots,k\bigr\}.
  \]
  A cell $C$ of the tree is a leaf if $V(C)$ meets at most one point class.  Otherwise $V(C)$
  contains two vertices $u,v$ in different point classes; they differ in some star coordinate
  $\X{1j}$, so $w(C)\ge\ext_{1j}(C)\ge\lvert\X{1j}(u)-\X{1j}(v)\rvert>0$.  Let $E=w(C)$, choose $j$
  with $\ext_{1j}(C)=E$ and a value $\lambda$ as in~\cref{cor:goodlevel}, and cut $C$ at
  $(1,j,\lambda)$.
\end{definition}

Since $\theta>0$, the chosen $\lambda$ satisfies $\min_C\X{1j}<\lambda<\max_C\X{1j}$, as required
in~\cref{def:hierarchy}.  The root contains every vertex point and, being convex, every curve
$\gamma_i$.

\begin{lemma}[termination]\label{lem:termination}
  The tree of~\cref{def:construction} is finite.
\end{lemma}

\begin{proof}
  Extents do not increase from a cell to a child, since the child is a subset.  Let $C$ be cut at
  $(1,j,\lambda)$ with $E=w(C)$.  By~\cref{lem:extent} the range $[m,M]$ of $\X{1j}$ on $C$ is the
  interval between the two bounds of $C$ in $\X{1j}$, and the children are the cells in which the
  upper, respectively the lower, of these bounds is replaced by $\lambda$; so
  $\ext_{1j}(C^-)=\lambda-m$ and $\ext_{1j}(C^+)=M-\lambda$.  As $M-m=E$ and $\lambda$ lies in the
  middle interval $[m+\theta E,M-\theta E]$ of~\cref{cor:goodlevel}, both are at most $(1-\theta)E$.

  \emph{Claim:} if $C'$ is a descendant of $C$ at distance $K=k-1$ below $C$, then
  $w(C')\le(1-\theta)w(C)$.  Suppose not.  Then every cell $C''$ on the branch from $C$ to $C'$
  has $w(C'')>(1-\theta)w(C)$, so each of the $K$ cuts on the branch is made in a star coordinate
  whose extent is $w(C'')>(1-\theta)w(C)$ before the cut and at most
  $(1-\theta)w(C'')\le(1-\theta)w(C)$ after it, and therefore for ever below.  Hence the $K$ cuts
  use $K$ different star coordinates, all $k-1$ star coordinates are used, and every star
  coordinate has extent at most $(1-\theta)w(C)$ in $C'$; that is, $w(C')\le(1-\theta)w(C)$, a
  contradiction.

  So the width decreases at least by the factor $1-\theta$ every $K$ levels along every branch.
  If all vertices lie in one point class, the root is a leaf.  Otherwise let $\delta>0$ be the
  smallest positive value of $\lvert\X{1j}(u)-\X{1j}(v)\rvert$ over all $j=2,\dots,k$ and all
  vertices $u,v$.  A cell $C$ with $w(C)<\delta$ is a leaf: for $u,v\in V(C)$ and every $j$,
  $\lvert\X{1j}(u)-\X{1j}(v)\rvert\le\ext_{1j}(C)\le w(C)<\delta$, so this difference is $0$; hence
  $u$ and $v$ agree in all star coordinates and lie in one point class.  Therefore every branch has
  length at most $K\bigl(\log(w(C_0)/\delta)/\log\frac{1}{1-\theta}+1\bigr)$, and the binary tree is
  finite.
\end{proof}

\subsection{Few pieces}

\begin{lemma}[pieces]\label{lem:pieces}
  For every cell $C$ of the tree of~\cref{def:construction}, the sets $\gamma_i^{-1}(C)$,
  $i=1,\dots,k$, have at most $\Pk$ connected components in total.  The paths $R_1,\dots,R_k$ have
  at most as many pieces in $V(C)$.
\end{lemma}

\begin{proof}
  Let $C_0\supset C_1\supset\dots\supset C_d=C$ be the branch from the root to $C$, where
  $C_{s+1}=C_s\cap H_s$ and $H_s$ is the closed half-space $\{\X{1j_s}\le\lambda_s\}$ or
  $\{\X{1j_s}\ge\lambda_s\}$ of the cut of $C_s$.  So $C=C_0\cap H_0\cap\dots\cap H_{d-1}$.

  Call
  the index $j_s$ together with the sign $\le$ or $\ge$ the \emph{type} of $H_s$; there are
  $2(k-1)$ types.  If $s<s'$ and $H_s,H_{s'}$ have the same type, say $\{\X{1j}\le\cdot\}$, then
  $\lambda_{s'}<\max_{C_{s'}}\X{1j}\le\lambda_s$, by the condition on the level in~\cref{def:hierarchy} and because $C_{s'}\subseteq H_s$; so $H_{s'}\subseteq H_s$.  The same holds
  for the sign $\ge$.  (In terms of bounds: by~\cref{lem:extent}, $\max_{C_{s'}}\X{1j}$ is the upper
  bound of $C_{s'}$ in $\X{1j}$, and every cut replaces one bound of the cell by its level; so the
  bounds of $C$ are those of $C_0$, except that each bound of a type that occurs is the level of
  the last half-space of that type.)  Hence $C=C_0\cap\bigcap_{H\in B}H$, where $B$ consists of the
  last half-space of each type that occurs, and $\lvert B\rvert\le 2(k-1)$.

  For $H=H_s\in B$ write
  $A_H=C_s$ (the cell that was cut by the hyperplane of $H$), $j_H=j_s$ and
  $\lambda_H=\lambda_s$; then $C\subseteq A_H$, and $N_{A_H}(\lambda_H)\le 3k-4$ with respect to
  $\X{1j_H}$ by the choice of the level.  The charging argument below uses only these facts,
  together with two more: every curve $\gamma_i$ lies in $C_0$, and every $\lambda_H$ is generic
  for $\X{1j_H}$.

  Fix $i$.  Since $\gamma_i$ is piecewise linear and $C$ is a convex polyhedron,
  $\gamma_i^{-1}(C)$ is a finite union of closed intervals; let $[t_1,t_2]$ be one of its
  components ($t_1=t_2$ is possible).

  Suppose $t_1>0$.  Then $\gamma_i(t)\notin C$ for all
  $t<t_1$ close to $t_1$, and on such an interval $(t_1-\varepsilon,t_1)$ the curve is affine.
  Every point $\gamma_i(t)$ lies in $C_0$, so it violates some $H\in B$.  For a given $H$, the set of
  $t$ in $(t_1-\varepsilon,t_1)$ where $H$ is violated is an interval, since an open half-space
  meets a segment in an interval.  Finitely many such intervals cover $(t_1-\varepsilon,t_1)$, so
  one of them contains $(t_1-\varepsilon',t_1)$ for some $\varepsilon'>0$.  Thus some $H\in B$ is violated at
  all $t<t_1$ close to $t_1$, and by continuity $\gamma_i(t_1)$ lies on the boundary hyperplane of
  $H$, $\{\X{1j_H}=\lambda_H\}$.  As $\gamma_i(t_1)\in C\subseteq A_H$, the pair $(i,t_1)$ is one
  of the at most $3k-4$ pairs counted by $N_{A_H}(\lambda_H)$.

  We \emph{charge} the left end of
  the component to the triple $(H,i,t_1)$.  In the same way, if $t_2<r_i$, we charge the right end
  to a triple $(H',i,t_2)$ where $H'\in B$ is violated at all $t>t_2$ close to $t_2$.

  The charging is injective.  Left ends of different components have different pairs $(i,t)$, and
  so do right ends.  If a left end and a right end are charged to the same triple $(H,i,t)$, then
  they belong to the same component $[t,t]$, and $H$ is violated at all points close to $t$ on both
  sides.  But $\lambda_H$ is generic, so $t$ is not a vertex parameter, $\gamma_i$ is affine near
  $t$, and the affine function $\X{1j_H}\circ\gamma_i$, which equals $\lambda_H$ at $t$, cannot lie
  strictly on the same side of $\lambda_H$ just before and just after~$t$.

  Hence at most $\sum_{H\in B}N_{A_H}(\lambda_H)\le 2(k-1)(3k-4)$ component ends are charged.  The
  ends that are not charged have $t_1=0$ or $t_2=r_i$; there are at most $2k$ of them.  Every
  component has two ends (equal if $t_1=t_2$, and then counted once as a left and once as a right
  end), so the number of components is at most $\bigl(2(k-1)(3k-4)+2k\bigr)/2=\Pk$.

  Finally, if $v^i_q$ and $v^i_{q+1}$ both lie in $V(C)$, the segment between their points lies
  in $C$, so consecutive vertices of a piece lie in one component, and a piece lies in one
  component.  Two different pieces of $R_i$ are separated by a vertex outside $C$, so they lie in
  different components.  Hence there are at most as many pieces as components.
\end{proof}

\begin{proof}[Proof of~\cref{thm:hierarchy}]
  Take the tree of~\cref{def:construction}.  It is a hierarchy and finite by~\cref{lem:termination}.  (H1) holds by the choice of $C_0$,~\ref{H:cut} by the choice of the
  levels (\cref{cor:goodlevel}),~\ref{H:pieces} by~\cref{lem:pieces}, and~\ref{H:leaf} by the
  leaf rule of~\cref{def:construction}.
\end{proof}

The bounds $3k-4$ and $\Pk$ depend on $k$ only.  The depth of the hierarchy depends on the
ratios of the values of the star coordinates (through $\delta$ and $w(C_0)$), but the algorithm of the next
section never constructs the hierarchy, and its running time does not depend on the depth.

\section{The algorithm}\label{sec:algorithm}

\subsection{Box sets}

For each $j=2,\dots,k$ let $\Lambda_j=\{\X{1j}(v):v\in V\}$ be the set of values of the star
coordinate $\X{1j}$ at vertices; $\lvert\Lambda_j\rvert\le n$.

\begin{definition}[box sets]\label{def:box}
  A \emph{box set} is a non-empty set of vertices of the form
  $U=\{v\in V:\alpha_j\le\X{1j}(v)\le\beta_j\text{ for }j=2,\dots,k\}$ with
  $\alpha_j,\beta_j\in\Lambda_j$.  Let $\Fam$ be the family of box sets.
\end{definition}

\begin{lemma}\label{lem:box}
  \leavevmode
  \begin{enumerate}[label=(\alph*)]
    \item $\lvert\Fam\rvert\le n^{2(k-1)}$, and $V\in\Fam$.
    \item A box set $U$ is determined by its \emph{tight bounds}
      $\min_{v\in U}\X{1j}(v)$ and $\max_{v\in U}\X{1j}(v)$, $j=2,\dots,k$: it is the box set with
      these bounds.
    \item If $C$ is a cell and $V(C)\neq\emptyset$, then $V(C)\in\Fam$.
    \item If $U\in\Fam$, $2\le j\le k$ and $\alpha\in\Lambda_j$, then
      $U\cap\{v:\X{1j}(v)\le\alpha\}$ and $U\cap\{v:\X{1j}(v)\ge\alpha\}$ are in $\Fam$ if they
      are non-empty.
    \item Every box set is a union of point classes.  A box set that meets only one point class
      is that point class.
  \end{enumerate}
\end{lemma}

\begin{proof}
  (a) There are $\lvert\Lambda_j\rvert^2\le n^2$ choices of $(\alpha_j,\beta_j)$ for each of the
  $k-1$ star coordinates.  $V$ is the box set with $\alpha_j=\min\Lambda_j$ and
  $\beta_j=\max\Lambda_j$.

  (b) Let $U$ have bounds $\alpha_j,\beta_j$ and tight bounds $\alpha'_j,\beta'_j$.  Then
  $\alpha_j\le\alpha'_j\le\beta'_j\le\beta_j$, so the box set $U'$ with the tight bounds satisfies
  $U\subseteq U'\subseteq U$.

  (c) Let $C$ have bounds $a_j,b_j$ (\cref{def:cell}).  For each $j$, $V(C)\ne\emptyset$ implies
  that $[a_j,b_j]\cap\Lambda_j\neq\emptyset$; replace $a_j$ by the smallest and $b_j$ by the largest
  element of this intersection.  This does not change the set of vertices satisfying the bounds,
  which is $V(C)$.

  (d) Replace $\beta_j$ by $\min(\beta_j,\alpha)$, or $\alpha_j$ by $\max(\alpha_j,\alpha)$.

  (e) Membership in a box set depends only on the star coordinates $\X{1j}(v)$, which are equal for
  all vertices of a point class.  So a box set is a union of point classes, and if it meets only
  one point class, it is that point class.
\end{proof}

By (b), box sets can be indexed by tuples of ranks of their tight bounds in the sorted lists
$\Lambda_j$, so that a box set given by its vertices or by its tight bounds is found in an
array in time depending on $k$ only after $O(kn)$ work to compute the bounds.

\subsection{Demands and the table}

A \emph{demand} is a triple $(i,a,b)$ with $i\in\{1,\dots,k\}$ (its \emph{label}) and
$a,b\in V$.  For $U\subseteq V$ let $D_i[U]$ be the set of arcs of $D_i$ with both ends in $U$.

\begin{definition}\label{def:S}
  For $U\subseteq V$ and a finite set $\Pi$ of demands, $\Sem(U,\Pi)$ is the statement: there are
  pairwise vertex-disjoint paths $R_\delta$, $\delta\in\Pi$, such that for $\delta=(i,a,b)$ the path
  $R_\delta$ leads from $a$ to $b$ and all its vertices lie in $U$ and all its arcs in $D_i[U]$ (for
  $a=b$, $R_\delta$ is the one-vertex path $a$).
\end{definition}

With $\Pi_0=\{(i,s_i,t_i):i=1,\dots,k\}$,~\cref{lem:Di}(b) shows that the instance has a solution if
and only if $\Sem(V,\Pi_0)$ holds.

\begin{definition}[the table]\label{def:table}
  The table has an entry $\Tab(U,\Pi)\in\{\text{true},\text{false}\}$ for every $U\in\Fam$ and
  every set $\Pi$ of at most $\Pk$ demands whose ends all lie in $U$.  The entries are computed in
  order of increasing $\lvert U\rvert$ by the following two rules.
  \begin{description}
    \item[Leaf rule.]  If $U$ is a point class: $\Tab(U,\Pi)=\Sem(U,\Pi)$ if no two demands of
      $\Pi$ have the same label, and $\Tab(U,\Pi)=\text{false}$ otherwise.  $\Sem(U,\Pi)$ is
      decided by~\cref{lem:pebbling} below.
    \item[Split rule.]  If $U$ meets at least two point classes: $\Tab(U,\Pi)$ is true if and only
      if there are a split and a valid interface (defined next) such that
      $\Tab(U^-,\Pi^-)$ and $\Tab(U^+,\Pi^+)$ are both true.
  \end{description}
  A \emph{split} $(j,\alpha,\alpha')$ of $U$ is an index $2\le j\le k$ together with two
  consecutive elements $\alpha<\alpha'$ of $\{\X{1j}(v):v\in U\}$; it defines
  $U^-=\{v\in U:\X{1j}(v)\le\alpha\}$ and $U^+=\{v\in U:\X{1j}(v)\ge\alpha'\}$.  Both are non-empty, they partition $U$, they are box sets
  (\cref{lem:box}(d)), and they are smaller than~$U$.

  An \emph{interface} for a split is a set $I$ of at most $3k-4$ pairs $(e,\delta)$, where
  $\delta=(i,a,b)\in\Pi$ and $e\in D_i$ is an arc with one end in $U^-$ and the other in $U^+$.  For
  a demand $\delta=(i,a,b)\in\Pi$ let $e_1,\dots,e_q$ be the arcs assigned to $\delta$ by $I$, sorted
  by the value of $F_i$ at their tails (if two tails have the same value, the interface is
  invalid), and write $e_s=(b_{s-1},a_s)$ and $a_0=a$, $b_q=b$.  This cuts $\delta$ into the
  \emph{chain} of \emph{sub-demands} $(i,a_0,b_0),(i,a_1,b_1),\dots,(i,a_q,b_q)$; for a demand with
  no assigned arc, $q=0$ and the chain is $\delta$ itself.  The interface is \emph{valid} if
  \begin{enumerate}[label=\textup{(V\arabic*)}]
    \item for every sub-demand $(i,a_s,b_s)$, the vertices $a_s$ and $b_s$ lie on the same side
      ($U^-$ or $U^+$);
    \item the ends $a_s,b_s$ of all sub-demands of all demands of $\Pi$ are pairwise distinct
      vertices, except that $a_s=b_s$ is allowed within one sub-demand;
    \item each side receives at most $\Pk$ sub-demands.
  \end{enumerate}
  Then $\Pi^-$ and $\Pi^+$ are the sets of sub-demands on the sides $U^-$ and $U^+$.
\end{definition}

The algorithm computes the table and answers $\Tab(V,\Pi_0)$; note $\lvert\Pi_0\rvert=k\le\Pk$.

\subsection{The leaf rule: pebbling inside a point class}

\begin{lemma}\label{lem:pebbling}
  Let $U$ be a point class and $\Pi=\{(i_1,a_1,b_1),\dots,(i_p,a_p,b_p)\}$ a set of demands with
  ends in $U$.  Then $\Sem(U,\Pi)$ can be decided, and paths witnessing it can be found, in time
  $O(p\,n^{p+1})$ for $p\ge 1$ and $O(1)$ for $p=0$.
\end{lemma}

\begin{proof}
  If $p=0$, return the empty family of paths.  Assume $p\ge 1$.
  By \Wtwo, every arc of every $D_i[U]$ increases $F_1$ by $2\ell(u,v)>0$.  Fix a total order
  $\prec$ of $U$ that refines the order by $F_1$ (break ties arbitrarily); every arc of
  $H=D_1[U]\cup\dots\cup D_k[U]$ goes from a vertex to a $\prec$-larger one.  If the $a_r$ are not
  pairwise distinct, or the $b_r$ are not, $\Sem(U,\Pi)$ is false; assume they are.

  We play a game with $p$ pebbles.  A state is a $p$-tuple $(x_1,\dots,x_p)$ of pairwise distinct
  vertices of $U$; the start is $(a_1,\dots,a_p)$ and the goal is $(b_1,\dots,b_p)$.  Pebble $r$ is
  \emph{finished} if $x_r=b_r$; finished pebbles never move.  A move: let $r$ be the unfinished
  pebble whose position is $\prec$-smallest, and replace $x_r$ by a vertex $y$ with
  $(x_r,y)\in D_{i_r}[U]$ that is not occupied by another pebble.  We claim that the goal is
  reachable from the start if and only if $\Sem(U,\Pi)$ holds.

  If $\Sem(U,\Pi)$ holds with paths $R_1,\dots,R_p$, let every pebble walk along its own path.  The
  pebble $r$ chosen by the rule is unfinished, so it is not at the end of $R_r$.  Its next vertex
  on $R_r$ is not occupied, because the other pebbles are on their own paths, which are disjoint
  from $R_r$.  Every move advances one pebble along its path, so the goal is reached.

  Conversely, consider a sequence of moves from the start to the goal.  Pebble $r$ traces a walk
  $W_r$ from $a_r$ to $b_r$ with arcs in $D_{i_r}[U]$; it is a path because $\prec$ increases along
  it.  Suppose two pebbles $r\neq q$ both visit a vertex $w$, and $r$ visits it first.  Pebble $r$
  must leave $w$ at some move, since otherwise $q$ could never enter $w$; so $r$ is unfinished at
  $w$, and at the moment it leaves, $r$ is the unfinished pebble with the $\prec$-smallest position.
  At that moment every other pebble is either finished, and then stays at its position, which is
  not $w$, or unfinished at a position $\succ w$, and then it only moves to $\prec$-larger vertices.
  In both cases $q$ never reaches $w$ afterwards, a contradiction.  So the paths $W_r$ are pairwise
  disjoint and witness $\Sem(U,\Pi)$.

  The game graph has at most $n^p$ states and each state has at most $n$ moves, each checked in
  time $O(p)$; reachability, and a play to the goal (which gives the paths), is found by graph
  search in time $O(p\,n^{p+1})$.
\end{proof}

\subsection{Correctness}

\begin{lemma}[soundness]\label{lem:sound}
  For every entry, $\Tab(U,\Pi)$ implies $\Sem(U,\Pi)$.
\end{lemma}

\begin{proof}
  Induction on $\lvert U\rvert$.  For the leaf rule this holds by definition.  Suppose the split
  rule makes $\Tab(U,\Pi)$ true through a split $(U^-,U^+)$ and a valid interface $I$ with
  $\Tab(U^-,\Pi^-)$ and $\Tab(U^+,\Pi^+)$ true.  By induction there are pairwise disjoint paths for
  the sub-demands in $\Pi^-$ inside $U^-$, and pairwise disjoint paths for those in $\Pi^+$ inside
  $U^+$; by (V2) different sub-demands are different elements of $\Pi^-\cup\Pi^+$, and since $U^-$
  and $U^+$ are disjoint, all these paths are pairwise disjoint.

  For a demand $\delta=(i,a,b)$ with
  chain $(i,a_0,b_0),\dots,(i,a_q,b_q)$, concatenate the paths of its sub-demands with the arcs
  $e_s=(b_{s-1},a_s)$.  This is a walk from $a$ to $b$ whose arcs all lie in $D_i$ and have both ends
  in $U$, and it is a path because its sub-paths are pairwise disjoint.  The paths obtained for
  different demands are disjoint, because they consist of disjoint sub-paths.  So $\Sem(U,\Pi)$
  holds.
\end{proof}

Soundness uses no property of the potentials: the order of the arcs in a chain and the choice of
the splits only matter for completeness.

\begin{lemma}[completeness]\label{lem:complete}
  If the instance has a solution, then $\Tab(V,\Pi_0)$ is true.
\end{lemma}

\begin{proof}
  Let $P_1,\dots,P_k$ be a solution; by~\cref{lem:Di}(b), $P_i$ is a $D_i$-path from $s_i$ to $t_i$.
  Apply~\cref{thm:hierarchy} to $\mathcal R=(P_1,\dots,P_k)$.  For a cell $C$ of the hierarchy with
  $V(C)\ne\emptyset$, let $\Pi(C)$ be the set of demands $(i,a,b)$, one for each piece of each $P_i$
  in $V(C)$, where $a$ and $b$ are the first and the last vertex of the piece.  The pieces are
  pairwise disjoint subpaths of the solution, so their demands are distinct, and
  $\lvert\Pi(C)\rvert\le\Pk$ by~\ref{H:pieces}; also $V(C)\in\Fam$ by~\cref{lem:box}(c).  So
  $(V(C),\Pi(C))$ is an entry of the table.  We show by induction from the leaves of the (finite)
  hierarchy that $\Tab(V(C),\Pi(C))$ is true for every cell $C$ with $V(C)\ne\emptyset$.  For the
  root, $V(C_0)=V$ and every $P_i$ is a single piece, so $\Pi(C_0)=\Pi_0$, which proves the lemma.

  \emph{Leaves.}  By~\ref{H:leaf}, $V(C)$ meets only one point class $Q$, so $V(C)=Q$ by~\cref{lem:box}(e) and the leaf rule applies.  By~\cref{lem:classpiece} every $P_i$ has at most
  one piece in $Q$, so no two demands of $\Pi(C)$ have the same label.  The pieces themselves
  witness $\Sem(V(C),\Pi(C))$: they are pairwise disjoint, the piece for $(i,a,b)$ leads from $a$ to
  $b$, and its vertices lie in $V(C)$ and its arcs in $D_i$.  So $\Tab(V(C),\Pi(C))$ is true.

  \emph{Cuts.}  Let $C$ be cut at $(1,j,\lambda)$ into $C^-$ and $C^+$, and write $U=V(C)$,
  $U^-=V(C^-)$, $U^+=V(C^+)$, so that $U$ is the disjoint union of $U^-$ and $U^+$.  If one of them
  is empty, the other child has the same vertex set as $C$ and the same pieces, so the same entry,
  which is true by induction.  Assume both are non-empty.  Since $C$ is not a leaf, $U$ meets at
  least two point classes~\ref{H:leaf}, so the table uses the split rule for $U$.  Let $\alpha$ be
  the largest value of $\X{1j}$ on $U^-$ and $\alpha'$ the smallest on $U^+$.  Since $\lambda$ is
  generic, $\alpha<\lambda<\alpha'$, these are consecutive values of $\X{1j}$ on $U$, and the split
  $(j,\alpha,\alpha')$ of the table produces exactly $U^-$ and $U^+$.

  Let $I$ be the set of pairs $(e,\delta)$ where $e$ is an arc of some $P_i$ with both ends in $U$,
  one in $U^-$ and one in $U^+$, and $\delta\in\Pi(C)$ is the demand of the piece of $P_i$ in $U$
  that contains $e$.  By~\cref{cor:crossarcs}, $\lvert I\rvert\le 3k-4$, and every such $e$ lies in
  $D_i$; so $I$ is an interface.

  Consider a piece of $P_i$ in $U$ with demand $\delta=(i,a,b)$.
  Along the piece $F_i$ increases strictly (by $2\ell(u,v)>0$ on every arc, \Wzero), so the arcs
  assigned to $\delta$, sorted by $F_i$ of their tails, appear in this order along the piece, and
  no two tails have the same value.  They cut the piece into maximal runs of consecutive vertices
  that lie alternately in $U^-$ and $U^+$, and the chain of $\delta$ consists of the demands of
  these runs.  A piece of $P_i$ in $U^-$ is contained in a piece of $P_i$ in $U$ (as
  $U^-\subseteq U$), and it ends where the piece in $U$ ends or where an arc of $I$ leaves $U^-$;
  so the runs are exactly the pieces of the paths in $U^-$ and in $U^+$, and $\Pi^-=\Pi(C^-)$,
  $\Pi^+=\Pi(C^+)$.

  (V1) holds because every run lies on one side.  (V2) holds because the ends
  of the runs are vertices of pairwise disjoint simple paths, at different positions unless they
  are the two ends of a one-vertex run.  (V3) holds by~\ref{H:pieces} for $C^-$ and $C^+$.  So $I$
  is valid, and $\Tab(U^-,\Pi(C^-))$ and $\Tab(U^+,\Pi(C^+))$ are true by induction; hence
  $\Tab(U,\Pi(C))$ is true.
\end{proof}

\begin{corollary}\label{cor:correct}
  $\Tab(V,\Pi_0)$ is true if and only if the instance has a solution.
\end{corollary}

\begin{proof}
  If $\Tab(V,\Pi_0)$ is true, then $\Sem(V,\Pi_0)$ holds by~\cref{lem:sound}, and the paths are
  shortest by~\cref{lem:Di}(b).  The converse is~\cref{lem:complete}.
\end{proof}

To output a solution, store with every true entry the split and the interface that made it true
(at a point class: the paths found by~\cref{lem:pebbling}), and unfold the construction of~\cref{lem:sound} from the root.

\subsection{Running time and proof of the main theorem}

\begin{lemma}\label{lem:time}
  After the preprocessing described in the proof, the table can be computed in time
  $f(k)\,\smash[t]{n^{6k^2-4k-1}}$, for a function $f$ of $k$ alone.
\end{lemma}

\begin{proof}
  Recall that $m\le n^2$ (no parallel arcs) and $\Pk=3k^2-6k+4$.

  \emph{Preprocessing.}  Compute $\dist(s_i,\cdot)$ and $\dist(\cdot,t_i)$ for all $i$ by $2k$ runs
  of Dijkstra's algorithm, in time $O(k(m+n\log n))$; then the sets $D_i$, the potentials, the
  star coordinates $\X{1j}(v)$ and the sorted lists $\Lambda_j$, in time $O(kn\log n+km)$.
  Enumerate all at most $n^{2(k-1)}$ bound tuples, compute their box sets and tight bounds, and
  store every box set once in an array indexed by its tight bounds (\cref{lem:box}(b)), in time
  $O(kn^{2k-1})$.  For every box set $U$ and every split of $U$ (at most $(k-1)n$ of them) record
  the indices of $U^-$ and $U^+$, in time $O(k^2n^{2k})$.  These two steps are within the bound
  of the lemma, since $2k\le6k^2-4k-1$ for $k\ge2$.

  Number the at most $kn^2$ demands; a set of at most
  $\Pk$ demands is stored as the sorted tuple of its numbers, so that, together with the index of
  $U$, every entry has an array position computable in time depending on $k$ only.

  \emph{Entries.}  There are at most $n^{2(k-1)}$ box sets and at most $(kn^2+1)^{\Pk}$ sets of at
  most $\Pk$ demands, so at most $n^{2(k-1)}(kn^2+1)^{\Pk}$ entries.

  \emph{Split rule.}  For one entry there are at most $(k-1)n$ splits.  The number of
  interfaces (sets of at most $3k-4$ pairs (arc, demand)) is at most
  $(m\Pk+1)^{3k-4}\le(n^2\Pk+1)^{3k-4}$.
  For each, building and sorting the chains, checking (V1)--(V3) and looking up the two entries
  takes time depending on $k$ only.  In total the split rule costs
  \[
    f_1(k)\,n^{2(k-1)}\cdot n^{2\Pk}\cdot n\cdot n^{2(3k-4)}
    =f_1(k)\,n^{2k-2+6k^2-12k+8+1+6k-8}=f_1(k)\,n^{6k^2-4k-1}.
  \]
  Up to factors depending on $k$, here $n^{2(k-1)}\cdot n^{2\Pk}$ bounds the entries, $n$ the
  splits and $n^{2(3k-4)}$ the interfaces, so the exponent is $2(k-1)+2\Pk+1+2(3k-4)$; for $k=3$
  (where $\Pk=13$) it is $41$.

  \emph{Leaf rule.}  At most $n$ box sets are point classes.  For each, at most
  $(n^2+1)^k\le2^kn^{2k}$ sets of demands with distinct labels need~\cref{lem:pebbling}, at time
  $O(kn^{k+1})$ each; all other entries at point classes are false and cost time depending on $k$
  only, and they are among the entries counted above.  Total
  $O(2^kk\,n^{3k+2})+f_2(k)\,n^{2(k-1)+2\Pk}$, which is within the bound because
  $3k+2\le6k^2-4k-1$ and $2(k-1)+2\Pk<6k^2-4k-1$ for $k\ge2$.
\end{proof}

\begin{proof}[Proof of~\cref{thm:main}]
  Remove parallel arcs in time $O(m+n)$ by bucketing the arcs by their ends.  If some $d_i$ is
  infinite, answer no.  Otherwise compute the table (\cref{lem:time}) and answer $\Tab(V,\Pi_0)$,
  which is correct by~\cref{cor:correct}; if it is true, unfold the stored choices to output a
  solution.  The shortest-path part of the preprocessing (distances, the sets $D_i$, potentials,
  star coordinates and the sorted lists $\Lambda_j$) takes $O(k(m+n\log n))$ time, which is
  $O(m+n\log n)$ for fixed $k$; the rest of the preprocessing and the table take
  $f(k)\,\smash[t]{n^{6k^2-4k-1}}$ time by~\cref{lem:time}.
  All numbers handled by the algorithm are distances, potentials and coordinates, that is,
  sums and differences of $O(n)$ arc lengths, and the algorithm only adds and compares
  them; the levels $\lambda$ of~\cref{sec:hierarchy} are never computed.
\end{proof}

\begin{remark}[two pairs]
  For $k=2$ the potential-difference space is a line with the single coordinate $\X{12}$, which
  increases along $P_1$ and decreases along $P_2$.  Cells are intervals, every generic level is
  crossed at most twice, $\Pk=4$, and the table is a divide-and-conquer version of a sweep over
  the values of $\X{12}$ with a two-pebble game inside each point class.  This is in the spirit of
  the algorithm of B\'erczi and Kobayashi~\cite{BercziKobayashi17}, who order pair $1$ forwards
  and pair $2$ backwards; we have not compared the two algorithms in detail.
\end{remark}

\section{Remarks and open problems}\label{sec:remarks}

\paragraph{Zero-length arcs.}  The main result is stated for positive lengths only.  Positive
lengths are used in two places: the chains of the split rule are ordered by $F_i$, which must
increase strictly along $D_i$, and inside a point class the union of the arc sets must be acyclic
(\cref{lem:pebbling}).  The following remark is not part of the main result, and it has not been
checked with the same care as the rest of the paper.

\begin{remark}\label{rem:zero}
  If the lengths are non-negative and $G$ has no directed cycle of length $0$, the lengths can be
  made positive without changing the sets of shortest $s_i$--$t_i$ paths.  The arcs of length $0$
  form an acyclic digraph; let $p(v)$ be the largest number of arcs of a path of zero-length arcs
  ending at $v$, let $\mu$ be the smallest positive arc length (or $1$ if there are no positive
  arc lengths), and put
  $\ell'(u,v)=\ell(u,v)+\frac{\mu}{n}(p(v)-p(u))$.  An arc of length $0$ has $p(v)\ge p(u)+1$, so
  $\ell'(u,v)\ge\mu/n$, and any other arc has $\ell'(u,v)\ge\mu-\mu(n-1)/n=\mu/n$.  On every path
  from $s$ to $t$ the added terms telescope to $\frac{\mu}{n}(p(t)-p(s))$, which depends only on
  $s$ and $t$.  So the algorithm of~\cref{thm:main}, run on $(G,\ell')$, would then decide this
  case as well.
\end{remark}

If zero-length cycles are allowed, the problem is NP-hard already for $k=2$~\cite{FHW80}: with all lengths $0$ every path is shortest, and the problem
contains the directed two disjoint paths problem.  In our framework all vertices then fall into
one point class and the leaf problem is the general disjoint paths problem.  For undirected
graphs, a zero-length edge gives a zero-length cycle of the corresponding digraph, so the remark
does not cover undirected graphs with zero-length edges; for $k=2$ these are solved in~\cite{GKW19,KobayashiSako19}.

\paragraph{What the proof uses.}  \Cref{thm:hierarchy} uses only \Wone, that is, that along the
$i$-th curve the coordinates $\X{ij}$ do not decrease.  It therefore holds for any $k$ polygonal
curves in $\R^k$ with this property, through any finite point set.

\paragraph{Arc-disjoint paths in directed graphs.}  The arc-disjoint version reduces to the vertex-disjoint one
through the line digraph, with the shortest-path condition preserved.

\begin{proposition}\label{prop:arcdisjoint}
  Let $G=(V,A)$ be a digraph, possibly with parallel arcs, with lengths $\ell\colon A\to\R_{>0}$,
  and let $(s_1,t_1),\dots,(s_k,t_k)$ be pairs with $s_i\neq t_i$, where terminals of different
  pairs may coincide.  Let $L$ be the digraph with a vertex $x_a$ for every arc $a\in A$ and
  $2k$ further vertices $\sigma_1,\tau_1,\dots,\sigma_k,\tau_k$, and with the arcs
  \begin{itemize}
    \item $(x_a,x_b)$ of length $\ell(b)$, whenever $a\neq b$ and the head of $a$ is the tail of $b$;
    \item $(\sigma_i,x_a)$ of length $\ell(a)$, whenever $s_i$ is the tail of $a$;
    \item $(x_a,\tau_i)$ of length $1$, whenever $t_i$ is the head of $a$.
  \end{itemize}
  Then $G$ has pairwise arc-disjoint paths $P_1,\dots,P_k$, each $P_i$ a shortest $s_i$--$t_i$ path
  of $G$, if and only if $L$ has pairwise vertex-disjoint paths $Q_1,\dots,Q_k$, each $Q_i$ a
  shortest $\sigma_i$--$\tau_i$ path of $L$.  The lengths of $L$ are positive, its $2k$ terminals
  are distinct, it has $\lvert A\rvert+2k$ vertices, and it is acyclic if $G$ is.
\end{proposition}

\begin{proof}
  The vertices $\sigma_j$ have no entering arcs and the $\tau_j$ no leaving arcs, so they are not
  inner vertices of any path.  Hence a $\sigma_i$--$\tau_i$ path of $L$ has the form
  $\sigma_i,x_{a_1},\dots,x_{a_r},\tau_i$ with $r\ge1$, where $a_1,\dots,a_r$ is an $s_i$--$t_i$
  walk of $G$ with pairwise distinct arcs, and every such walk gives such a path; its length is
  $\ell(a_1)+\dots+\ell(a_r)+1$.  Since the lengths are positive, a shortest $s_i$--$t_i$ walk of $G$
  is a path, and every path is a walk with distinct arcs.  So the distance from $\sigma_i$ to
  $\tau_i$ in $L$ is $\dist(s_i,t_i)+1$, and the shortest $\sigma_i$--$\tau_i$ paths of $L$ are
  exactly the images of the shortest $s_i$--$t_i$ paths of $G$.

  Two such images are
  vertex-disjoint if and only if the paths are arc-disjoint, because their inner vertices are the
  arcs of the paths and their end vertices are distinct and not inner vertices.  The lengths of $L$
  are positive by construction.  A directed cycle of $L$ avoids the $\sigma_j$ and $\tau_j$, so it
  gives a closed walk of $G$; hence $L$ is acyclic if $G$ is.
\end{proof}

\paragraph{Edge-disjoint paths in undirected graphs.}  Replacing every edge by two opposite arcs
does not reduce the undirected edge-disjoint version to the arc-disjoint one, since two paths may
then traverse one edge in opposite directions.  For example, the graph with a single edge $uv$
and the pairs $(u,v)$ and $(v,u)$ has no edge-disjoint solution, but its two opposite arcs give
an arc-disjoint one.  The line graph, with a single vertex for both directions of an edge, reduces it to the
directed vertex-disjoint version instead.

\begin{proposition}\label{prop:edgedisjoint}
  Let $G=(V,E)$ be an undirected graph, possibly with parallel edges, with lengths
  $\ell\colon E\to\R_{>0}$, and let $(s_1,t_1),\dots,(s_k,t_k)$ be pairs with $s_i\neq t_i$, where
  terminals of different pairs may coincide.  Let $L$ be the digraph with a vertex $y_e$ for every
  edge $e\in E$ and $2k$ further vertices $\sigma_1,\tau_1,\dots,\sigma_k,\tau_k$, and with the arcs
  \begin{itemize}
    \item $(y_e,y_f)$ of length $\ell(e)+\ell(f)$, whenever $e\neq f$ have a common end;
    \item $(\sigma_i,y_e)$ of length $\ell(e)$, whenever $s_i$ is an end of $e$;
    \item $(y_e,\tau_i)$ of length $\ell(e)$, whenever $t_i$ is an end of $e$.
  \end{itemize}
  Then $G$ has pairwise edge-disjoint paths $P_1,\dots,P_k$, each $P_i$ a shortest $s_i$--$t_i$ path
  of $G$, if and only if $L$ has pairwise vertex-disjoint paths $Q_1,\dots,Q_k$, each $Q_i$ a
  shortest $\sigma_i$--$\tau_i$ path of $L$.  The lengths of $L$ are positive, its $2k$ terminals
  are distinct, and it has $\lvert E\rvert+2k$ vertices.
\end{proposition}

\begin{proof}
  As before, the $\sigma_j$ and $\tau_j$ are not inner vertices of any path, so a
  $\sigma_i$--$\tau_i$ path $Q$ of $L$ has the form $\sigma_i,y_{e_1},\dots,y_{e_r},\tau_i$ with
  $r\ge1$ and pairwise distinct edges $e_1,\dots,e_r$, where $s_i\in e_1$, $t_i\in e_r$, and
  $e_j,e_{j+1}$ share an end for every $j<r$.  Let $H_Q$ be the subgraph of $G$ formed by these
  edges, and $\ell(H_Q)=\ell(e_1)+\dots+\ell(e_r)$.  The arcs of $Q$ have the lengths $\ell(e_1)$,
  $\ell(e_1)+\ell(e_2)$, \dots, $\ell(e_{r-1})+\ell(e_r)$, $\ell(e_r)$, so $Q$ has length
  $2\ell(H_Q)$.  Since $H_Q$ is connected and contains $s_i$ and $t_i$, it contains an
  $s_i$--$t_i$ path $P$, and $\dist(s_i,t_i)\le\ell(P)\le\ell(H_Q)$.  Conversely, an
  $s_i$--$t_i$ path of $G$ with edges $e_1,\dots,e_r$ in this order gives the path
  $\sigma_i,y_{e_1},\dots,y_{e_r},\tau_i$ of $L$, its \emph{image}, of twice its length.  So the
  distance from $\sigma_i$ to $\tau_i$ in $L$ is $2\dist(s_i,t_i)$, and the image of a shortest
  path is a shortest path.

  If $P_1,\dots,P_k$ are as in the statement, their images are pairwise vertex-disjoint, because
  their inner vertices are the vertices $y_e$ of the edges of the paths and their end vertices are
  distinct and not inner vertices.  Conversely, let $Q_1,\dots,Q_k$ be as in the statement, and
  let $P_i$ be an $s_i$--$t_i$ path in $H_{Q_i}$.  Since $Q_i$ is shortest,
  $\ell(H_{Q_i})=\dist(s_i,t_i)$, so $\ell(P_i)\le\dist(s_i,t_i)$ and $P_i$ is a shortest path;
  and the $P_i$ are pairwise edge-disjoint because the $Q_i$ are pairwise vertex-disjoint.  The
  lengths of $L$ are positive by construction.
\end{proof}

(We checked the proposition by brute force on
all multigraphs with at most $4$ vertices, at most $5$ edges, lengths in $\{1,2\}$ and two pairs,
and on $36000$ random multigraphs with up to $7$ vertices, $9$ edges and $4$ pairs, many with
parallel edges and coinciding terminals; no disagreement.)

The two propositions give the corollary stated in the introduction.

\begin{restated}{cor:edgedisjoint}
  \coredgedisjoint
\end{restated}

\begin{proof}
  In the undirected case, loops lie on no path.  Between two vertices only edges of minimum
  length lie on shortest paths, and a path uses at most one edge between them.  So deleting the
  loops, keeping $k$ such edges (all, if there are fewer) between any two vertices and deleting
  the others changes neither the distances nor the answer.  This takes linear time and leaves at
  most $k\binom n2$ edges.  In
  the directed case the same trimming, applied to the arcs from $u$ to $v$ for every ordered pair
  $(u,v)$, with loops deleted, leaves at most $k\,n(n-1)$ arcs.

  Then~\cref{prop:arcdisjoint}, in the
  directed case, and~\cref{prop:edgedisjoint}, in the undirected case, give in polynomial
  time an equivalent instance of directed vertex-disjoint $k$-DSPP.  It has positive lengths,
  $2k$ distinct terminals and at most $k n^2+2k$ vertices, and~\cref{thm:main} solves it in time
  $\smash[t]{n^{O(k^2)}}$ for fixed~$k$.
\end{proof}

\paragraph{Routes through prescribed vertices.}  In digraphs, deciding whether there is a path
from $s$ to $t$ through a vertex $v$, or a cycle through two vertices $s$ and $t$, is
NP-complete~\cite{FHW80}.  In undirected graphs, a shortest cycle through $k$ prescribed vertices
can be found in randomized time $2^k n^{O(1)}$~\cite{BHT12}.  With polynomially bounded integer
edge weights, a shortest walk through $O(\log n)$ waypoints, visited in any order, can be found
in randomized polynomial time~\cite{AFS20}.

A route through $w_0,\dots,w_\ell$ in this order has length at least
$L=\sum_{i=1}^{\ell}\dist(w_{i-1},w_i)$.  Whether this bound is attained is a disjoint shortest
paths problem with $\ell$ pairs, so it can be decided in polynomial time for every fixed $\ell$,
in directed graphs as well.

\begin{corollary}\label{cor:waypoints}
  Let $\ell\ge 2$ be fixed.  Let $G$ be a directed or undirected graph with $n$ vertices, $m$
  arcs or edges and strictly positive lengths, let $w_0,\dots,w_\ell$ be vertices of $G$, and
  let $L=\sum_{i=1}^{\ell}\dist(w_{i-1},w_i)$.  For each of the following kinds of route, one can
  decide in time $O(m)+\smash[t]{n^{O(\ell^2)}}$ whether $G$ has a route of this kind of length
  $L$, and find one if it exists.
  \begin{enumerate}[label=(\roman*)]
    \item A path from $w_0$ to $w_\ell$ that passes through $w_1,\dots,w_{\ell-1}$ in this order.
      Here $w_0,\dots,w_\ell$ are pairwise distinct.
    \item A cycle that passes through $w_0,\dots,w_{\ell-1}$ in this cyclic order.  Here
      $w_\ell=w_0$, the vertices $w_0,\dots,w_{\ell-1}$ are pairwise distinct, and $\ell\ge 3$ if $G$ is
      undirected.
    \item A walk from $w_0$ to $w_\ell$ that passes through $w_1,\dots,w_{\ell-1}$ in this order
      and uses every arc or edge at most once.  Here $w_{i-1}\ne w_i$ for every $i$.
  \end{enumerate}
  Every route of each kind has length at least $L$.
\end{corollary}

\begin{proof}
  We may assume that $L$ is finite.  Split a route of any of the three kinds at its visits to the
  prescribed vertices into segments $R_1,\dots,R_\ell$, where $R_i$ goes from $w_{i-1}$ to
  $w_i$.  The segment $R_i$ has length at least $\dist(w_{i-1},w_i)$, so the route has length at
  least $L$, with equality if and only if every $R_i$ has length $\dist(w_{i-1},w_i)$.  As the
  lengths are positive, such a segment is a shortest $w_{i-1}$--$w_i$ path.

  In case (iii), a walk of length $L$ is therefore the same as $\ell$ pairwise arc-disjoint, or
  edge-disjoint, shortest paths for the pairs $(w_{i-1},w_i)$.  Terminals of different pairs
  may coincide, and~\cref{cor:edgedisjoint} decides this.

  In cases (i) and (ii), if $G$ is undirected, replace every edge by two opposite arcs of the
  same length.  This changes neither the distances nor the paths, and since $\ell\ge3$ in case
  (ii), a directed cycle through the $\ell$ prescribed vertices is a cycle of $G$.  Let $W$ be the
  set of the prescribed vertices other than $w_0$ and $w_\ell$ in case (i), and the set
  $\{w_0,\dots,w_{\ell-1}\}$ in case (ii).  Form $G'$ by replacing every $w\in W$ by two vertices:
  $w^-$, which receives the arcs into $w$, and $w^+$, which sends the arcs out of $w$.  For
  $w\notin W$ put $w^-=w^+=w$.  The pair $i$ has the source $w_{i-1}^+$ and the target $w_i^-$;
  these $2\ell$ terminals are distinct.

  A vertex $w^-$ has no outgoing arcs and $w^+$ has no incoming arcs, so a path of $G'$ passes
  through such a vertex only at its ends.  Hence the paths of $G'$ from $w_{i-1}^+$ to $w_i^-$
  are the $w_{i-1}$--$w_i$ paths of $G$ with no inner vertex in $W$.  A route of kind (i) or (ii)
  is the same as $\ell$ such paths, one for each pair, that are pairwise vertex-disjoint in $G'$.
  Let $\dist'$ be the distance in $G'$.  If $\dist'(w_{i-1}^+,w_i^-)>\dist(w_{i-1},w_i)$ for
  some $i$, no route has length $L$, since the segment $R_i$ has no inner vertex in $W$.
  Otherwise, the shortest paths of $G'$ for the pair $i$ are the shortest $w_{i-1}$--$w_i$ paths
  of $G$ with no inner vertex in $W$.  Then a route of length $L$ exists if and only if $G'$ has
  pairwise vertex-disjoint shortest paths for the $\ell$ pairs.  \Cref{thm:main} decides this,
  since $G'$ has at most $n+\ell$ vertices.
\end{proof}

If the order of the prescribed vertices is not given, trying all orders decides whether the
shortest route through them has length $\min_\sigma L_\sigma$, the smallest of the bounds for
the orders $\sigma$.  When the bound is attained, \cref{cor:waypoints} also finds a shortest route.

\paragraph{The exponent.}  Our bound $\smash[t]{n^{6k^2-4k-1}}$ is not optimized.  Undirected $k$-DSPP with
unit lengths is W[1]-hard parameterized by $k$~\cite{BNRZ21}, and so is the directed problem on
DAGs with unit lengths~\cite{AW21}.  So, unless FPT${}={}$W[1], the exponent cannot be bounded
independently of $k$.

The exponent is quadratic in $k$ only because of the number of pieces per cell,
$\Pk=3k^2-6k+4$.  A cell has $2(k-1)$ walls, each crossed at most $3k-4$ times, and each piece
contributes two end vertices to a table entry.  All other terms of the exponent are linear in $k$.  Cutting along
all $\binom k2$ coordinates $\X{ij}$ instead of the $k-1$ coordinates $\X{1j}$ lowers the number
of crossings per cut to $2k-2$, but gives cells with up to $k(k-1)$ walls and a cubic exponent.
Whether the exponent can be
brought down to $O(k)$, or to $O(k\log k)$, for example with $O(k)$ pieces per cell, is open.

\paragraph{Open problems.}
\begin{itemize}
  \item \emph{Faster algorithms.}  For $k=2$ much faster algorithms are known~\cite{CKS26,CQW26}.  Is there an algorithm for fixed $k\ge3$ with a small explicit exponent,
    or a practical implementation of the guillotine table (for example bottom-up over box sets
    that actually contain demands)?
  \item \emph{Lower bounds for fixed $k$.}  What conditional lower bounds can be proved for
    directed $k$-DSPP with positive lengths for small fixed values of $k$, in particular $k=3$?
\end{itemize}

\section*{Formal verification}
\addcontentsline{toc}{section}{Formal verification}
\label{sec:formal}

The main results of this paper, \cref{thm:main}, \cref{cor:undirected}, \cref{cor:edgedisjoint},
\cref{prop:arcdisjoint}, \cref{prop:edgedisjoint} and~\cref{cor:waypoints}, are formalized in
Lean~4 with Mathlib, together with all lemmas used in their proofs.\footnote{\url{https://github.com/saeedamiri/disjoint-shortest-paths-lean}}
The formalization has no unproved statements and uses only the standard axioms of Lean.  It also
proves that the paths read off the table form a solution.  The running time is formalized as a
count of the table's work; the time of the preprocessing is not formalized.  The file
\texttt{DEVIATIONS.md} in the repository lists the differences in presentation.

\section*{Declaration on the use of AI}
\addcontentsline{toc}{section}{Declaration on the use of AI}

The author provided the problem statement and the initial resources, including the literature on
disjoint shortest paths, the open question for directed graphs, and his own unpublished manuscripts.
The main ideas and the algorithm of this paper, in particular the potentials and the guillotine cut
hierarchy, were discovered with Claude Fable, an AI model developed by Anthropic.  The author
verified all proofs.  The proofs were also checked by independent reviews carried out by AI agents and by
computational tests.  The research, the formalization and the reviews were carried out with
an agent harness created by the author.\footnote{\url{https://github.com/saeedamiri/agent-skills}}
The results are formalized in Lean~4, as described in the section
\hyperref[sec:formal]{Formal verification}.  The Lean code was written with AI assistance.
The author used AI assistance (GPT Sol and Claude Opus) in writing this paper and worked throughout
to simplify it.  The author takes full responsibility for its content.

\bibliographystyle{alphaurl}
\bibliography{refs}

\end{document}